\documentclass[aps,prd,superscriptaddress,nofootinbib]{revtex4-1}
\pdfoutput=1

\usepackage{booktabs}
\usepackage[english]{babel}
\usepackage{amsmath,amssymb,amsbsy,amstext, amsthm, simplewick,slashed}
\usepackage{graphicx}
\usepackage{amsfonts}
\usepackage{tikz}
\usepackage{upgreek}
\usepackage{simplewick}
\usepackage{exscale,relsize}
\usepackage{bbding}
\usepackage{tabu}
\usepackage{multirow}
\usepackage{color}
\usepackage{graphicx}  
\usepackage{dcolumn}   
\usepackage{bm}        
\usepackage{amssymb}   
\usepackage{amsmath} 
\usepackage{enumerate}
\usepackage{graphicx}  
\usepackage{dcolumn}   
\usepackage{bm}        
\usepackage{amssymb}   
\usepackage{amsmath} 
\usepackage{enumerate}

\newcommand{\jets}{\text{jets}}

\newcommand{\tev}{\text{TeV}}
\newcommand{\gev}{\text{GeV}}
\newcommand{\fbinv}{\text{fb}^{-1}}

\usepackage{tikz}
\usetikzlibrary{decorations.pathmorphing,decorations.markings}

\tikzset{
photon/.style={decorate, decoration={snake,amplitude=4pt, segment length=7pt}, draw=black},
particle/.style={draw=black, postaction={decorate}, decoration={markings,mark=at position .5 with {\arrow[draw=black]{>}}}},
antiparticle/.style={draw=black, postaction={decorate}, decoration={markings,mark=at position .5 with {\arrow[draw=black]{<}}}},
gluon/.style={decorate, draw=black, decoration={coil,amplitude=3pt, segment length=4pt}},
higgs/.style={draw=black,dashed,thick },
arrow/.style={draw=black, very thick, postaction={decorate}, decoration={markings,mark=at position 1 with {\arrow[draw=black]{>}}}}
}

\definecolor{darklightsabergreen}{rgb}{0.0, .49, 0.06}

\begin{document}

\thispagestyle{empty}

\vspace*{1.31cm}

\begin{center}
{\LARGE \bf {Cornering a Hyper Higgs: Angular Kinematics for Boosted Higgs Bosons with Top Pairs}}\\

\vspace*{1.61cm} {\large
Joseph Bramante\footnote{\tt
jbraman2@nd.edu},
Antonio Delgado\footnote{\tt antonio.delgado@nd.edu}, and
Adam Martin\footnote{\tt amarti41@nd.edu}}\\
\vspace{.5cm}
Department of Physics, University of Notre Dame, 225 Nieuwland Hall, Notre Dame, IN, USA, 46556\\

\vspace{3.5cm}

\end{center}

{In the wake of the Higgs discovery and over the long haul of the LHC run, one should keep a lookout for kinematic anomalies in the most massive known trio of coupled particles, $t \bar{t} h$. After surveying the scope of prior constraints on chromomagnetic dipole and Higgs-gluon kinetic couplings, we focus on surpluses of boosted-$p_T$ Higgs bosons fomented by these momentum dependent dimension-six operators in $t \bar{t} h$ final states. We uncover a number of simple, $p_T$ weighted angular variables useful for discriminating Standard Model from dimension-6 boosted Higgs distributions, and make headway arguing that one of these variables may improve the reach of existing Standard Model top-Higgs searches. The approach we take is model independent, because we just consider a set of effective operators that contribute to the same three-body final state.}
\vspace{8cm}

\setcounter{page}{0} \setcounter{footnote}{0}


\date{\today}




\section{Introduction}
Studies of the new scalar boson discovered by the Large Hadron Collider (LHC) thus far have evinced couplings matching those of the Standard Model (SM) Higgs boson. Because the Higgs boson is now the most obvious mechanism for giving masses to fermions, an accurate measurement of its Yukawa couplings will be of paramount importance in the coming decade.  The largest of these Yukawa couplings, the Higgs-top coupling, will be the most illuminated when the LHC takes data at $\sqrt{s}= 14$ TeV. Hence it is timely to propose new experimental methods and non-standard physics for the Higgs-top Yukawa coupling, both for application to existing $\sqrt{s}= 7-8$ TeV data and future 14 TeV LHC events.

One model-independent approach is to study new high-scale physics through an effective field theory (EFT) analysis of $(n \geq 5)$-dimension Higgs couplings to Standard Model particles~\cite{Burges:1983zg, Leung:1984ni, Buchmuller:1985jz,Hagiwara:1993ck,Whisnant:1997qu, Grzadkowski:2010es}. The top quark Yukawa coupling is nearly unity. Thus, dimension-6 effective top-Higgs operators are the most incisive in discovering and limiting Yukawa-channel physics beyond the Standard Model. Recent effective operator analyses have shown how the rates observed in $h$ and $h + X$ final states at the LHC can constrain non-standard top-Higgs dimension-6 couplings \cite{Degrande:2012gr,Hayreter:2013kba}. 

However, as LHC observations of $h+X$ event rates converge on Standard Model values, it is essential to move beyond event rates and illuminate what unique kinematic distributions in $h + X$ final states could also signal new physics at higher energies. In this paper we study the novel kinematics of dimension six operators that produce more highly-boosted $p_T$ Higgs events than those found in Standard Model $t\bar{t}h$ final states. Specifically, the dimension-6 chromomagnetic dipole coupling and the Higgs-gluon kinetic coupling are examined, because their momentum structure leads to non-standard boosted $p_T$ Higgs bosons in $p p \rightarrow t \bar{t} h$ production. New kinematic variables and event selection strategies are developed to single out these new events versus the SM background.

In the next section we determine what bounds can be placed on Higgs boosting dimension-6 couplings using existing LHC studies. In Section \ref{sec:boostedpthiggs} we explain how these operators boost the Higgs in $t\bar{t}h$ events. Section \ref{sec:dileptonhighpt} develops a set of detector level kinematic variables useful both for finding $t\bar{t}h$ events with a boosted Higgs and for refining Standard Model $t \bar{t} h$ searches. These variables are implemented for multilepton $t \bar{t}h$ searches in Section \ref{section:cmscutsleptonic}, and we determine what variable combinations best reveal boosted Higgs and reduce the dominant backgrounds, which are $t \bar{t} W^\pm$ and $t \bar{t} + \jets$ with a jet faking a lepton. In Section \ref{section:cmscutsphotonic} a simpler photon $p_T$ sieve is found for boosted Higgs digamma decay in $t \bar{t}h$. Finally Section \ref{sec:conclusion} is devoted to our conclusions.

\section{Rates and distributions: Higgs-gluon and chromomagnetic dipole couplings}
\label{sec:ratemubounds}

When multiple new physics operators contribute to a process (such as $h+X$), the total rate does not uniquely constrain the size of the individual operators since there can be cancellations among the different contributions. One way to break the leftover degeneracy is to compare multiple processes that are influenced by the same new physics, i.e. $pp \to h$ and $pp \to t\bar{t}h$ are both sensitive to modifications in the top-quark Yukawa coupling. A second approach is to focus on a single process but study different kinematic regimes, exploiting the fact that new physics operators often have different kinematic structure than the SM counterparts. For example, an amplitude may have the form
\begin{equation}
\mathcal A = \mathcal A_{SM}  + c_{\mathcal O} \frac{q^2}{\Lambda^2}\mathcal A_{\mathcal O},
\label{eq:simple}
\end{equation}
where $\mathcal A_{SM}$ is the SM piece (taken to be momentum independent), $c_{\mathcal O}$ is a dimensionless coefficient accompanying a new physics operator $\mathcal O$, $\Lambda$ is the scale suppressing the new physics, and $q$ is a momentum transfer in the process. The presence of powers of $q$ is due to extra derivatives in the new-physics operator $\mathcal O$. From the form of Eq.~(\ref{eq:simple}) it is clear that the effect of $c_{\mathcal O}$ are enhanced if we look at high-momentum regions of phase space. Similarly, processes characterized by low momentum transfer will not be very sensitive to $c_{\mathcal O}$. These momentum-dependent new physics effects, in the context of $t\bar t h$ production, are the target of our study here. Similar studies of momentum-dependent effects have previously been performed in the $h + W/Z$ modes in Ref.~\cite{Ellis:2013ywa, Isidori:2013cga,Isidori:2013cla, Alloul:2013naa} and $h + j$ in Ref.~\cite{Banfi:2013yoa,Azatov:2013xha,Grojean:2013nya}. Other recent studies of $t\bar t h$ (or $t/\bar t + h$) that do not exploit momentum-dependent new physics can be found in Ref.~\cite{Curtin:2013zua,Craig:2013eta,Onyisi:2013gta} (Ref.~\cite{Englert:2014pja}). \\

While there are many Higgs effective operators whose cutoffs and couplings are set by the scale and dynamics of new high energy physics, only a small subset are relevant for $pp \to t\bar t h$ production. Of the dim-$6$ operators relevant for $t\bar t h$, only two have non-SM momentum structure: the dimension-6 chromomagnetic dipole and Higgs-gluon kinetic coupling. These two effective Higgs couplings arise naturally in many extensions of the Standard Model and result in an increased Higgs $p_T$ in $t \bar{t} h$ final states as compared to Standard Model distributions. Boosted Higgses provide unique kinematic signatures, and this is the focus of most of this study. However in the remainder of this section, we will first delineate bounds on dimension-6 Higgs-top and Higgs-gluon couplings using LHC studies of Higgs, $t \bar{t}$ and $t \bar{t}h$ events. As we will see, the twenty inverse femtobarn run of the LHC has already set substantial limits on non-standard Higgs couplings by bounding the production rates of $pp \to h$, $t \bar{t}$ and $t \bar{t}h$. 

Throughout this work we hew closely to the conventions of \cite{Degrande:2012gr}. We assume the cutoff of our effective operators is at the TeV scale, $\Lambda = 1 {~\rm TeV}$ and take the Higgs mass to be $m_H = 126 {~\rm GeV}$. To define our new physics operators, we specify a Standard Model Lagrangian supplemented by two effective field theory dimension-6 Higgs operators, 
\begin{align}
\mathcal{L} = \mathcal{L}_{SM} + \mathcal{O}_{hgt} +\mathcal{O}_{HG}. \label{eq:efflag}
\end{align} 
These dimension-6 operators are the Higgs-gluon kinetic operator 
\begin{align}
\mathcal{O}_{HG} = \frac{c_{HG}}{2\Lambda^2}\left(H^\dag H \right)G_a^{\mu \nu}G^a_{\mu \nu}. \label{eq:ohg}
\end{align}
and the chromomagnetic dipole term,
\begin{align}
\mathcal{O}_{hgt} = \frac{c_{hgt}}{\Lambda^2} \left(\bar{Q}_L H \right) \sigma^{\mu \nu} T^a t_R G^a_{\mu \nu}, \label{eq:ohgt}
\end{align}
where $ \sigma^{\mu \nu} = i\left[\gamma^\mu,\gamma^\nu\right]$ and ${\rm Tr}[T^aT^b]=\delta^{ab}/2$. There are other dimension-6 operators involving Higgses, top-quarks, and/or gluons that we can write down, such as
\begin{align}
\mathcal O_{cHq} = (\bar Q_{3L}\gamma^{\mu}\, Q_{3L})(H^{\dag}\overleftrightarrow{D_{\mu}}H),\quad & \mathcal O_{c'Hu} = (\bar Q_{3L}\sigma_i \gamma^{\mu}\, Q_{3L})(H^{\dag}\sigma^i\overleftrightarrow{D_{\mu}}H),\quad \mathcal O_{cHu} = (\bar t_{R}\gamma^{\mu}\, t_{R})(H^{\dag}\overleftrightarrow{D_{\mu}}H) \nonumber \\
&\text{and}\quad \mathcal O_{y_u} = H^{\dag}H\, \bar Q_{3L} H^c t_R.
\label{eq:extraops}
\end{align}

Here we are neglecting different 4-fermion operators that could possibly contribute to FCNC and which would therefore be very constrained. Moreover, we are only directly modifying the top sector, although this same approach could also be applied to other quarks, especially to the bottom.
The first three operators involve an electroweak boson and between zero to two Higgses, so they do not contribute to $t\bar t h$ production at leading order in $\alpha_s$. The final operator contains only tops and Higgses. Once two of the three Higgses are set to their vacuum expectation value, this operator adds to the the top-quark Yukawa coupling. While including $\mathcal O_{y_u}$ will certainly impact $pp \to h$ and $pp \to t\bar t h$ rates, this operator has the same momentum structure as the SM. Since our focus is on operators with non-SM kinematics, we will set the coefficient of $\mathcal O_{y_u}$ to zero for the rest of this work. This is just a simplification -- the techniques we outline in Sec.~\ref{sec:boostedpthiggs} will still work in the presence of $\mathcal O_{y_u}$.  Also, we ignore any possible CP-violating higher dimensional operators.

Returning to the operators in Eq.~(\ref{eq:efflag}), both terms affect the rate of $p p \rightarrow h$ production. The greater part of their contribution to Higgs production can be written as a modification of the Standard Model gluon fusion production rate. Specifically, it has been shown at leading order \cite{Degrande:2012gr,Manohar:2006gz} that in the heavy top limit $\mathcal{O}_{HG}$-associated Higgs production is related to SM Higgs production by
\begin{align*}
\frac{\sigma(pp\rightarrow h)}{\sigma(pp\rightarrow h)_{SM}} \equiv \mu_{h} \simeq \left( 1+ c_{HG}\frac{24 \pi^2  v^2}{\Lambda^2 g_s^2}\right)^2,
\end{align*}
where $v=246 {~\rm GeV}$ is the Higgs vacuum expectation value. Note that $\mathcal{O}_{HG}$ sources Higgs particles at tree level. As we will presently show, this tightly constrains the size of $(c_{HG}/ \Lambda^2)$ as compared to $(c_{hgt}/ \Lambda^2)$, which results in a smaller, one-loop contribution to $c_{HG}$ for $pp \rightarrow h$ production \cite{Degrande:2012gr},
\begin{align}
\delta c_{HG} = c_{hgt} \frac{g_s y_t\,\log\left(\frac{\Lambda^2}{m_t^2}\right)}{4 \pi^2}.
\end{align}
Here $m_t$ is the top mass.

\begin{figure}[h!]
\centering
\includegraphics[scale=1]{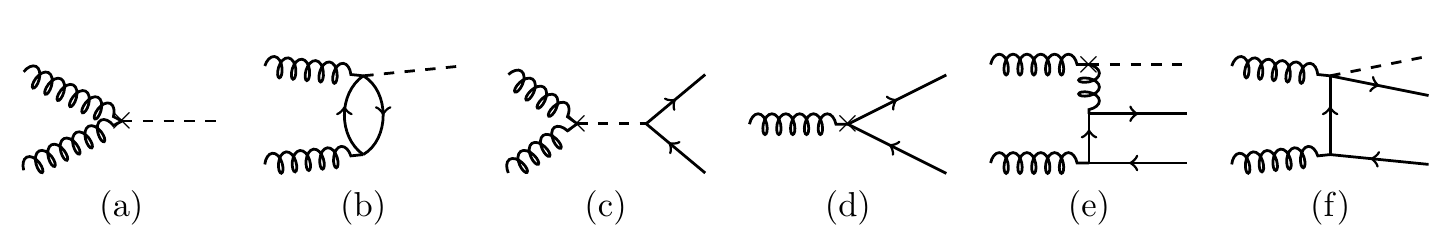}
\caption{The Feynman diagrams shown above exemplify non-standard Higgs, $t \bar{t}$, and $t \bar{t}h$ processes resulting from the chromomagnetic dipole and Higgs-gluon operators studied in this paper. Diagrams (a) and (b) show contributions to Higgs production from Higgs-gluon coupling (Eq. \ref{eq:ohg}) and the chromomagnetic dipole (Eq. \ref{eq:ohgt}), respectively. Top pair production is also affected by the Higgs-gluon kinetic coupling and chromomagnetic dipole as illustrated in (c) and (d). Finally, new processes producing $t \bar{t}h$ shown in (e) and (f) include vertices with the Higgs coupled to a gluon -- these vertices lead to a larger number of $t \bar{t}h$ events with boosted Higgs bosons.}
\label{fig:diagrams}
\end{figure}

\begin{figure}[h!]
\centering
\begin{tabular}{cc}
\includegraphics[scale=.55]{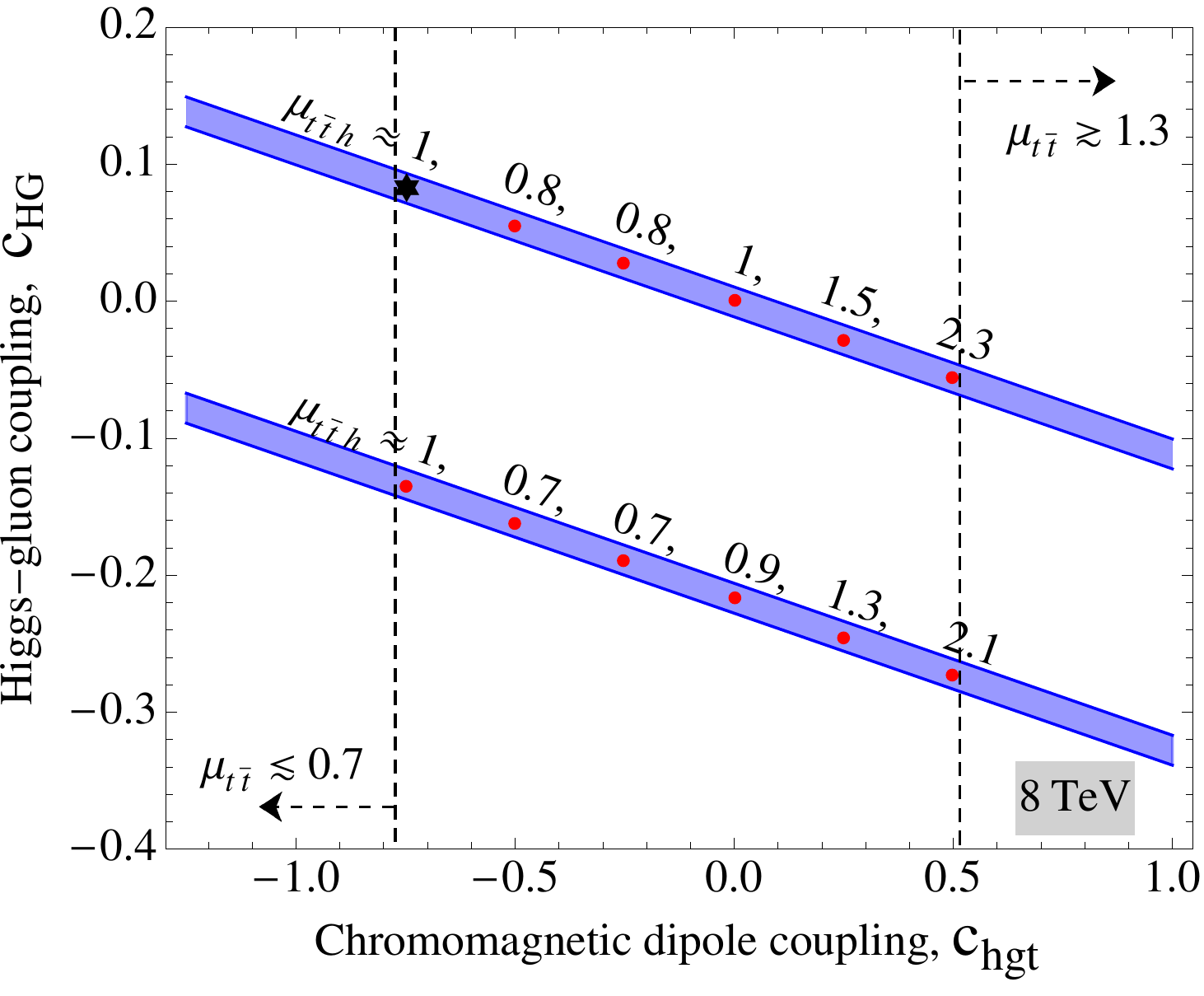} & \includegraphics[scale=.55]{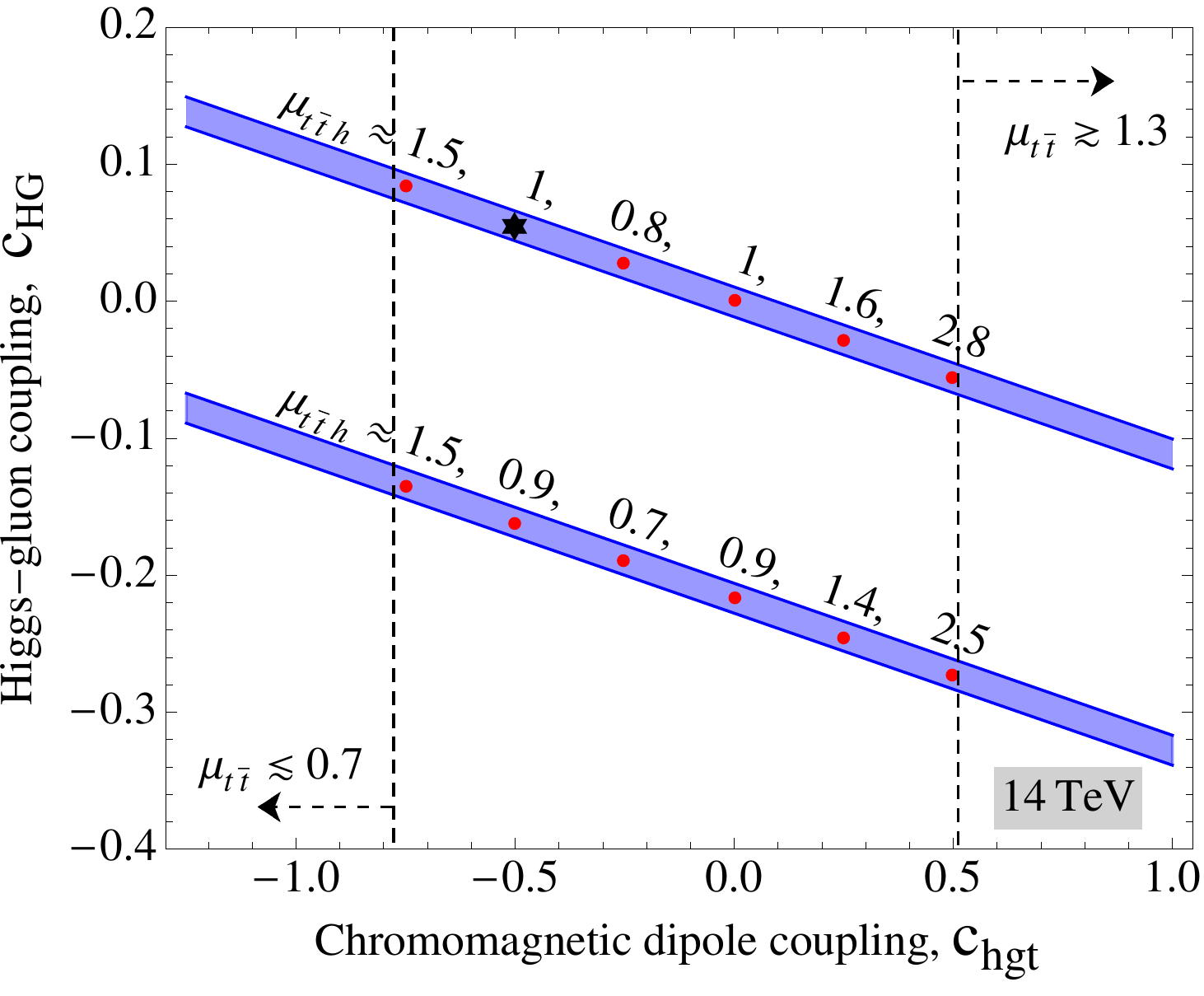}
\end{tabular}
\caption{Bounds on the dimension-6 chromomagnetic dipole moment and Higgs-gluon kinetic coupling are shown. The blue bands show parameter space allowed assuming that Higgs final state production has been limited to $0.8 < \mu_h < 1.2$. Twelve model points surveyed in this study are plotted, along with their predicted $\mu_{t\bar{t}h}$ values. Notice that the CMS combined limit of $\mu_{tth} < 4.4$ at 95\% does not constrain these operators more than measurements of the $p p \rightarrow t \bar{t}$ cross section. Bounds on $c_{hgt}$ are given with dashed lines, assuming at most a 30\% deviation from the SM $t \bar{t}$ expected rate, $0.7 < \mu_{t \bar{t}} < 1.3$. Note that the only difference between the 8 TeV and 14 TeV constraints plotted on the left and right are the $\mu_{tth}$ values.}
 \label{fig:feynpph}
\end{figure}

Studies of $pp \to h$, $t \bar{t}$ and $t \bar{t} h$ production at the LHC set limits on $\mathcal O_{HG}$ and $\mathcal O_{hgt}$. As shown diagrammatically in Fig.~\ref{fig:diagrams},  a chromomagnetic dipole will affect all these final states, and the Higgs-gluon kinetic coupling will alter $h$ and $t \bar{t} h$ at tree level. Indeed, in order to study large yet viable couplings for these operators, the contributions from these two operators to inclusive Higgs production must nearly cancel out\footnote{Had we included $\mathcal O_{y_u}$ (Eq.~(\ref{eq:extraops})) the cancellation among new physics contributions to $pp \to h$ would involve $c_{y_u}$ as well as $c_{HG}$ and $c_{hgt}$. See Ref.~\cite{Degrande:2012gr}.}. In Fig.~\ref{fig:feynpph} we show the constraints imposed on ($c_{HG}$, $c_{hgt}$) parameter space by LHC studies of $p p \rightarrow h$, $p p \rightarrow t \bar{t}$, and $p p \rightarrow t \bar{t}  h$.

A combination of Higgs production studies from the ATLAS collaboration finds $\mu_h = 1.43 \pm 0.21 ~(1\sigma)$ \cite{ATLAS-CONF-2013-014}, while the CMS combination finds $\mu_h = 0.87 \pm 0.23~(1\sigma)$ \cite{Chatrchyan:2012ufa}. In Fig.~\ref{fig:feynpph} we shade blue parameter space in the ($c_{hgt},\,c_{HG}$) plane that is consistent with a Higgs production of $\pm 20\%$ of the Standard Model expectation. Note that this is the part of the parameter space where new Higgs production contributions of the two dimension-6 operators cancel. The two bands correspond to the two different solutions of the quadratic equation that relates the coefficients of the new physics operators to the total cross-section.

The chromomagnetic dipole term in Eq.~\ref{eq:ohgt} will alter the  tree-level $t \bar{t}$ production, predominantly through s-channel gluon-associated production of top pairs, as illustrated in Fig.~\ref{fig:diagrams}(d). The new kinetic Higgs-gluon coupling will also alter $t \bar{t}$ production (Fig.~\ref{fig:diagrams}(c)), however because limits on $\mu_h$ constrain $|c_{HG}| \lesssim 0.2$, its affect is negligible. Both ATLAS and CMS have measured top pair production in a number of channels, and a combination of their results at 7 TeV \cite{ATLAS-CONF-2012-134-CMS-PAS-TOP-12-003} found $\sigma_{t \bar{t}} = 173.3 \pm 2.3_{\rm{stat.}} \pm 9.8_{\rm{syst.}} \rm{pb}$ in agreement with a Standard Model prediction of $\sigma_{t \bar{t}} = 167 \pm 18\, \rm{pb}$. In Fig.~\ref{fig:feynpph} dashed lines mark what values of $c_{hgt}$ are consistent with a top pair production rate within $\pm 30\%$ of the Standard Model rate.

Limits on $t \bar{t}h$ couplings are less stringent: a recent technical report on a combination of $t \bar{t} h$ studies from the CMS collaboration finds $\mu_{t \bar{t} h} < 4.3 ~(2\sigma)$ \cite{CMSTtHwiki}, while an ATLAS study in the Higgs decay channel $t \bar{t} h \rightarrow t \bar{t} \gamma\gamma$ limits $\mu_{t \bar{t} h} < 5.3 ~(2\sigma)$ \cite{ATLAS-CONF-2013-080}. These measurements should improve during the 14 TeV run with more luminosity. In Fig.~\ref{fig:feynpph} we label twelve points in parameter space with $\mu_{t \bar{t} h}$ ratios. It is clear that in the parameter space of interest, current $t \bar{t}$ studies exclude more phase space than $t \bar{t}h$ studies. While $t \bar{t}$ studies conducted thus far have set stringent limits on the Wilson coefficients $c_{HG}$ and $c_{hgt}$ taken alone, here we reemphasize that other new physics (Eq. \ref{eq:extraops}) contributions could cancel the contributions of these coefficients. This motivates kinematic $t \bar{t}h$ searches, even if the non-standard kinematic operators shown here are naively more constrained by other search channels (e.g. $t \bar{t} h$).

In quoting limits on e.g. $\sigma (pp \rightarrow t\bar{t}h \rightarrow t\bar{t} \gamma \gamma)$, $\sigma (pp \rightarrow t\bar{t}h \rightarrow t\bar{t} W^+ W^-)$, and using these to limit $pp \rightarrow t\bar{t}h$ production generally, we assume Standard Model Higgs branching ratios to $\gamma \gamma$, $ZZ$, $WW^*$, $b \bar{b}$, and $\ell \bar{\ell}$. This is a well-motivated supposition given that these ratios are thus far consistent with the Standard Model \cite{ATLAS-CONF-2013-014,Chatrchyan:2012ufa}. For studies on the effect of dimension-$6$ operators on Higgs decays, see~\cite{Han:2013sea}.
 
Finally, we emphasize that the bounds derived in this section are valid only for our setup of Eq.~(\ref{eq:efflag}), i.e. the SM extended by two particular operators. If more operators were added, the bounds would change, though not necessarily for all channels. For example, if the operator $\mathcal O_{qqtt} = (\bar q_{\lambda} \gamma^{\mu}\, q_{\lambda})(\bar t_{\lambda'} \gamma_{\mu}\, t_{\lambda'}), (\lambda, \lambda' = L,R)$ were added, the bounds on $pp \to t\bar t$ and $t \bar t h$ would change, while $pp \to h$ would not change (at least at lowest order). Similarly, adding the operator $\mathcal O_{y_u}$ would change $pp \to h$ and $pp \to t\bar t h$ while hardly affecting $t\bar t$ production. The points shown in Fig.~\ref {fig:feynpph} therefore should not be interpreted as strict bounds on $c_{HG}$ or $c_{hgt}$ -- rather they are a rough indication of what size effects are permitted with the current dataset.
 
\section{Boosted $p_T$ Higgs}
\label{sec:boostedpthiggs}

Prior studies of non-standard Higgs-top couplings focus on the observed rates of $t \bar{t} h$ at the LHC in a number of different final states: 
\begin{align}
t \bar{t} h \rightarrow t \bar{t} + \{b \bar{b}, ~\tau \bar{\tau}, ~W^+ W^-,~ZZ, ~ \gamma \gamma \}.
\end{align}
Studies of these Higgs decay channels have produced constraints on $\mu_{t\bar{t}h}$ for each Higgs decay, and have been combined in analyses like \cite{CMSTtHwiki}. These limits on the rate of $ p p \rightarrow t\bar{t}h$ can be applied to the effective operators $c_{HG}$ and $c_{hgt}$, as we have done in Section \ref{sec:ratemubounds} and Fig.~\ref{fig:feynpph}.

\begin{figure}[h!]
\centering
\includegraphics[scale=1]{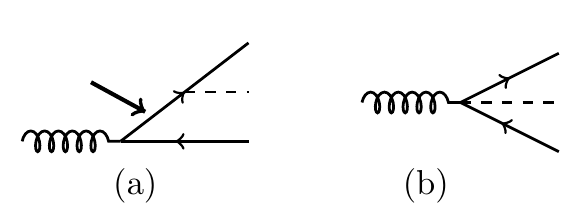}
\caption{Two s-channel processes demonstrate how the chromomagnetic dipole operator shifts Higgs $p_T$ to higher values in $t\bar{t} h$ events. The first (a) is a Standard Model process, which has an internal top propagator (indicated by an arrow) that is proportional to a factor of $\sim 1/P$ (where $P$ is the 4-momentum of the top and Higgs) as compared to the new physics process (b) which creates $t \bar{t} h$ at a single vertex with strength proportional to $(p_h + p_t + p_{\bar t})$.} 
\label{fig:boostdiag}
\end{figure}

The feature we explore in this study is that for $t \bar{t}h$ events produced by the operators $c_{HG}$ and $c_{hgt}$, the kinematic distribution of Higgs bosons in $t\bar{t}h$ final states will differ substantially from those produced in the Standard Model, even if the total $pp \to h$ and $pp \to t\bar t h$ rates are near SM-valued. Specifically, in the presence of $c_{HG}$ and $c_{hgt}$, the $t\bar t g$ vertex acquires a momentum dependence not present in the SM, and there is a new, momentum dependent, four particle ($t\bar t h g$) vertex. Following the argument of Eq.~(\ref{eq:simple}), the effects of this momentum-dependent new physics grow as the energy of the final state increases. The Higgs, being the lightest of the $t\bar t h$ trio, typically carries a larger transverse momentum than the $t$ or $\bar t$. When more energy is pumped into the $t\bar t h$ final state, this trend continues, and the transverse momentum distribution of the Higgs becomes increasingly skewed towards higher values. The two Feynman diagrams shown in Fig.~\ref{fig:boostdiag} highlight how the momentum-dependence enters into $s$-channel $t\bar t h$ amplitude.  The first diagram shows a SM configuration, where the vertices are momentum independent and high momentum Higgses come at the expense of putting the internal top quark propagator (indicated with an arrow) farther off-shell. The second diagram comes from the chromomagnetic dipole term -- there is no internal propagator and the $t\bar t h g$ vertex is proportional to the sum of the momenta of all three final state particles, leading to a strong preference for energetic $t\bar t h$ configurations. Similar arguments can be made for other diagrams in Fig.~\ref{fig:feynpph}. There is, of course, a limit to the amount of energy one can give to the $t\bar t h$ system which is given by $\Lambda$, the cut-off in our operator expansion, when the energy starts being close to 1 TeV the EFT breaks down and a suitable UV completion is needed. We will therefore only analyze situations where the energy of the boosted Higgs is below $\Lambda$. 

\begin{figure}[h!]
\centering
\includegraphics[scale=.6]{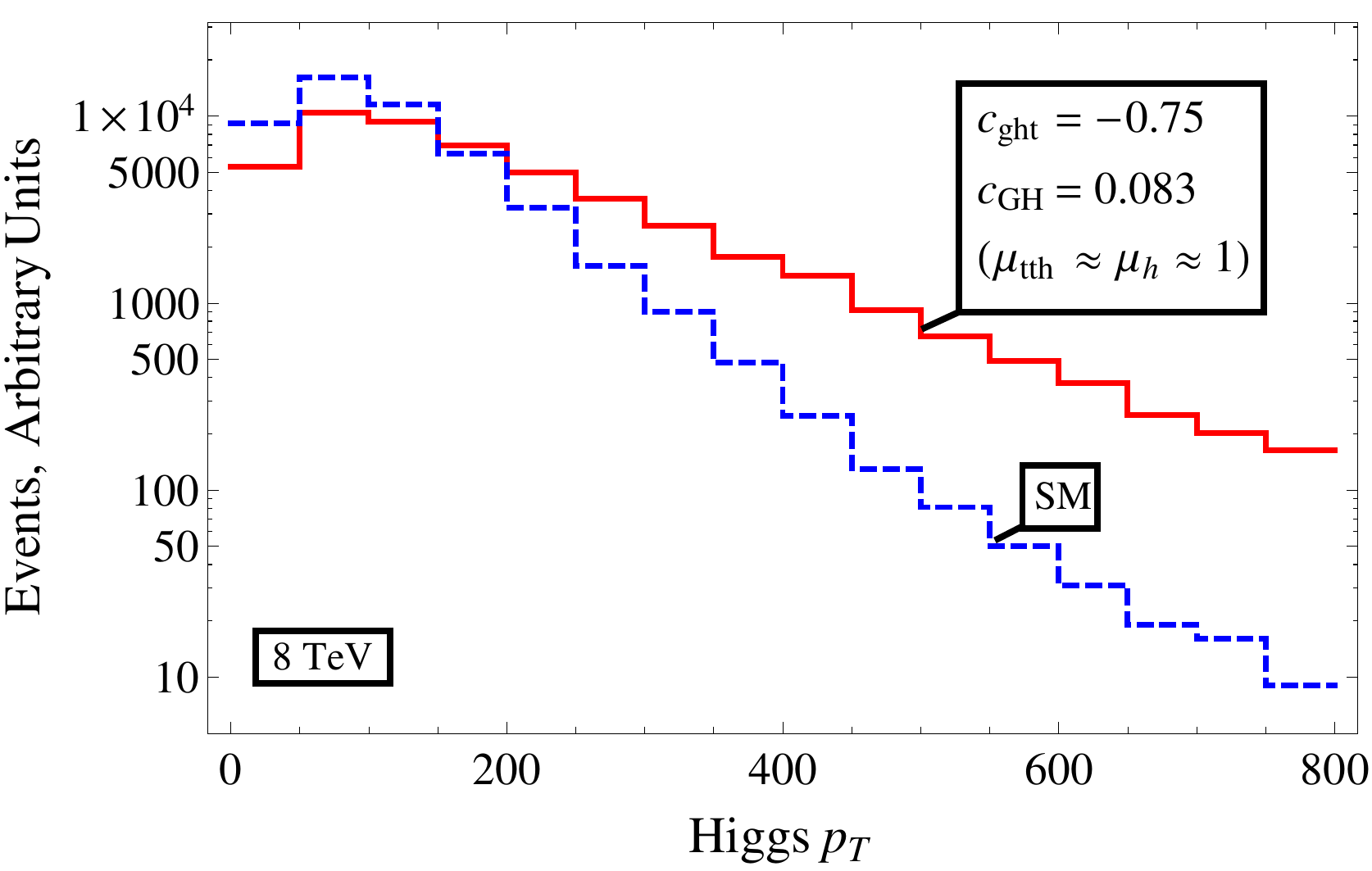}
\begin{tabular}{cc}
\\
\includegraphics[scale=.6]{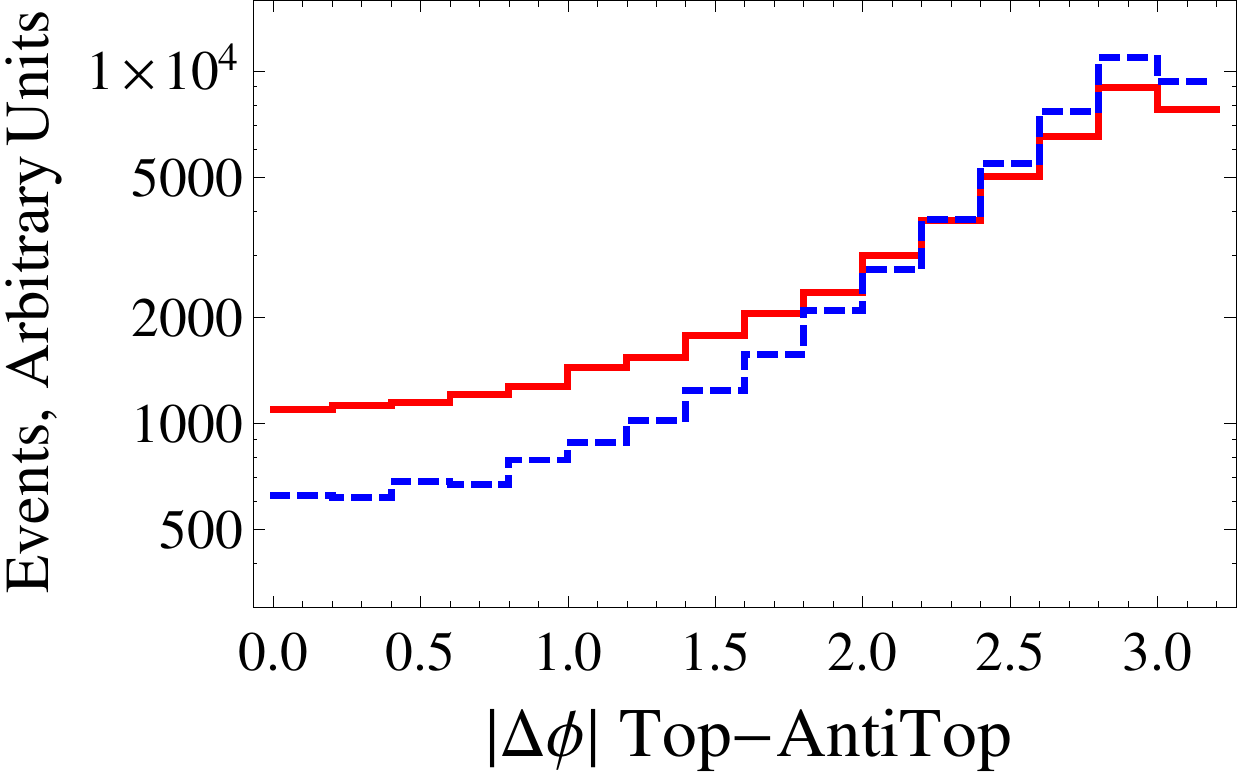}&\includegraphics[scale=.6]{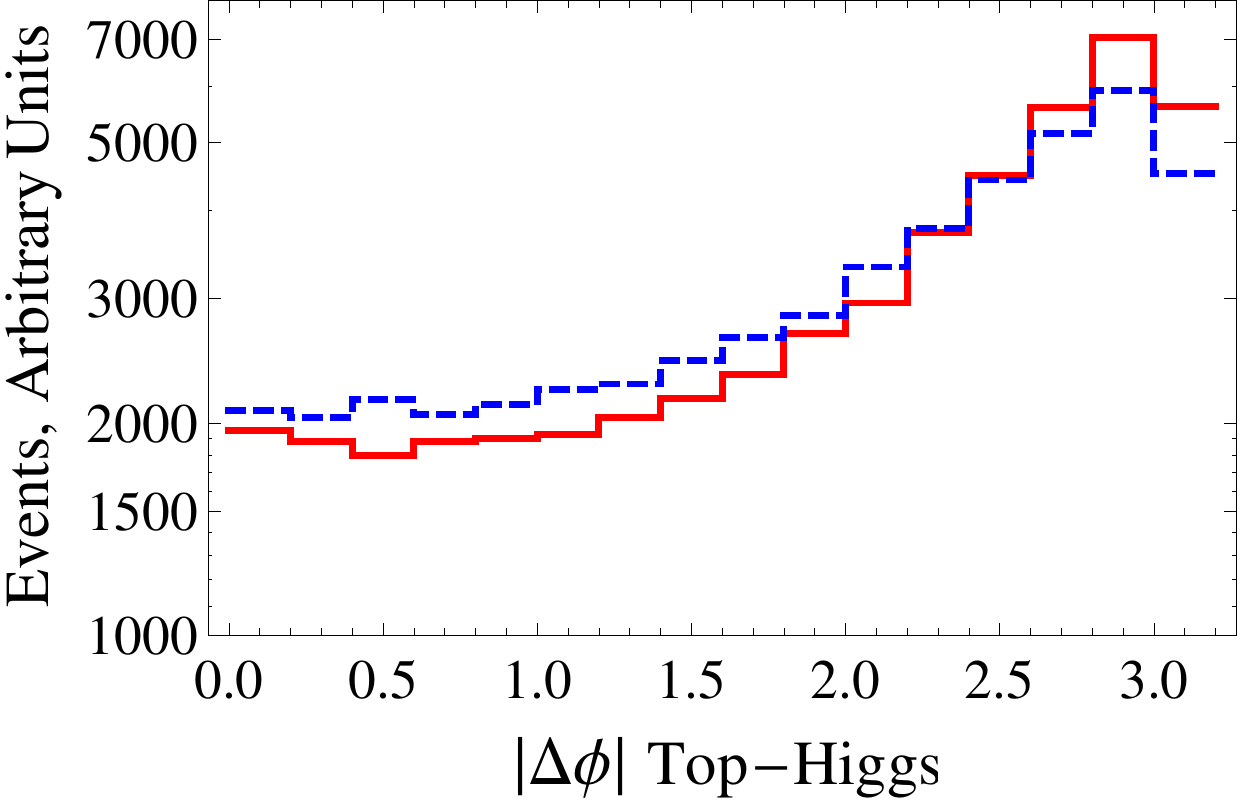} \\
\includegraphics[scale=.6]{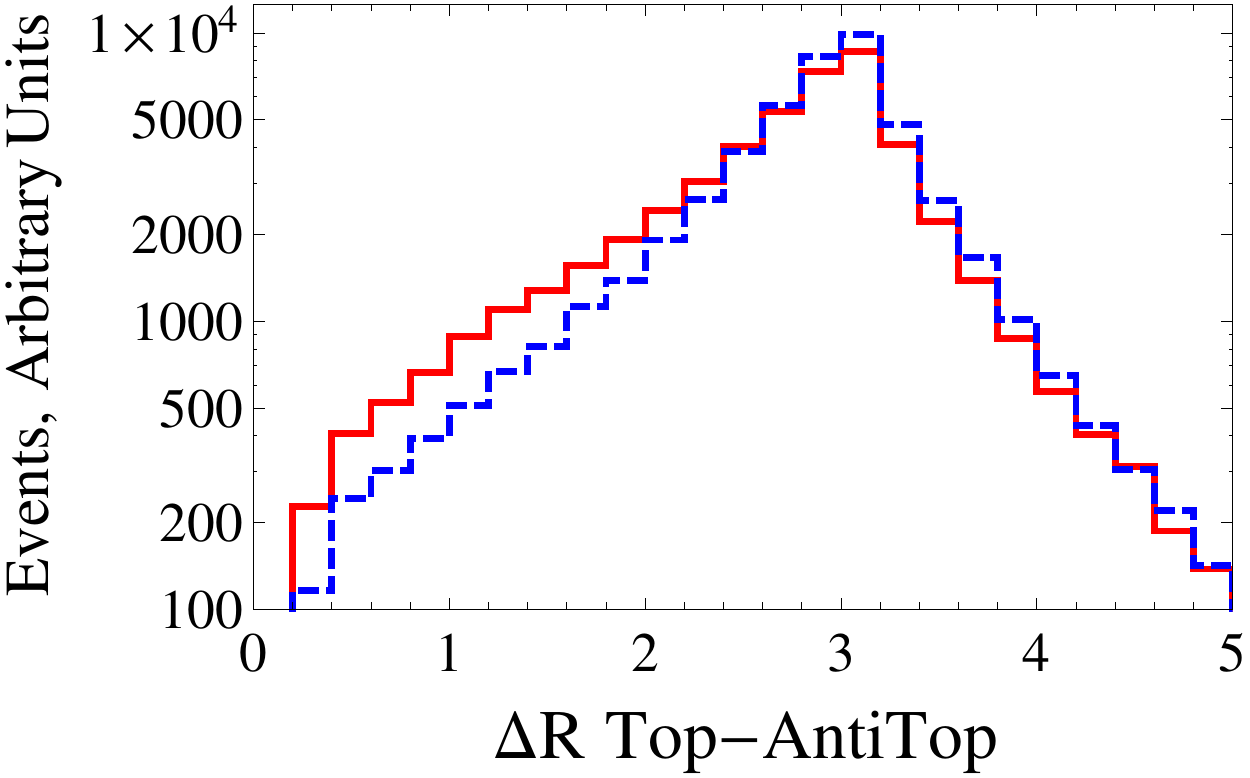}&\includegraphics[scale=.6]{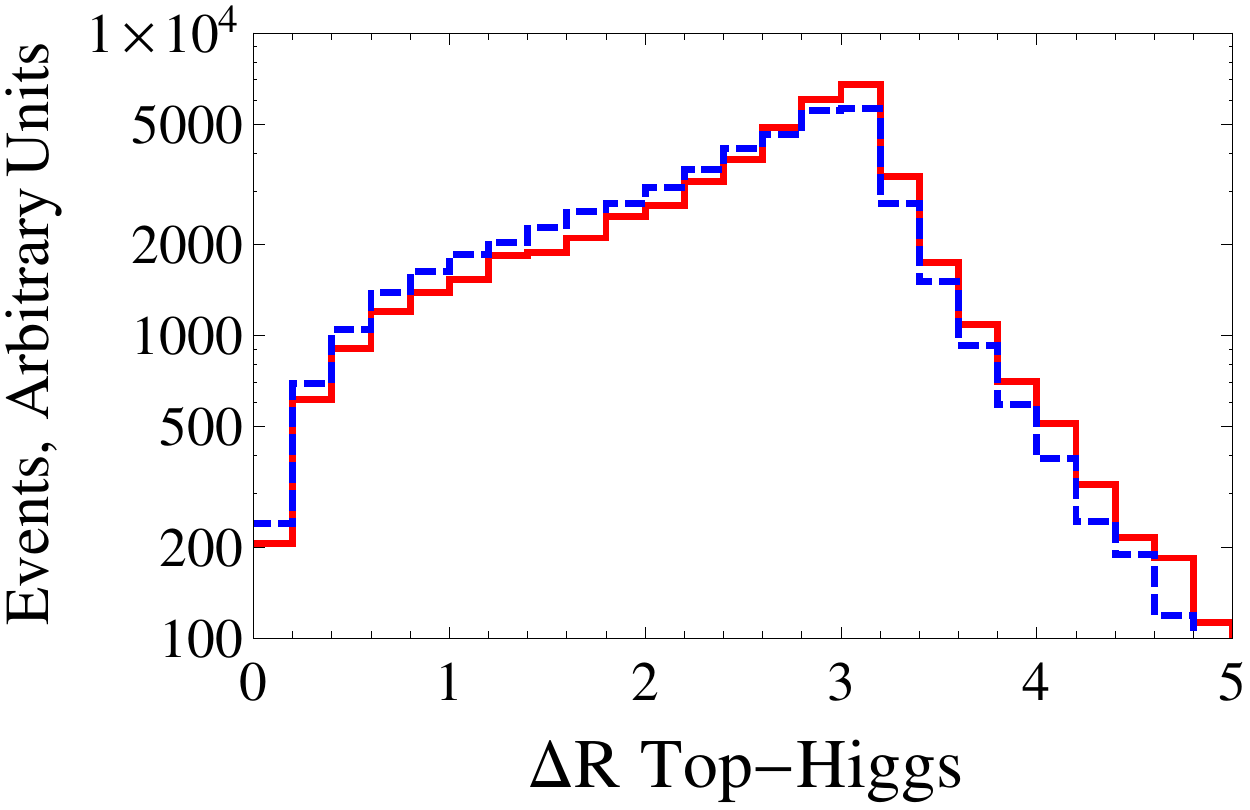} 
\end{tabular}
\caption{Kinematics distributions before cuts: In the upper plot, the transverse Higgs momentum is given for $t \bar{t}h$ final states at $\sqrt{s} = 8 ~ \rm{TeV}$ both in the SM (blue) and for a new physics scenario with dimension-6 effective operator coefficients $c_{hgt}=-0.75$ and $c_{HG}=0.083$~(red). The predicted $p p \rightarrow t \bar{t}h$ and $p p \rightarrow h$ cross-sections for the new physics scenario are nearly identical to the standard model -- the difference in Higgs transverse momenta is a result of non-standard kinematics. In the lower plots, the angular separation of partons for $t\bar{t}h$ events again for the SM and for $c_{hgt}=-0.75$ and $c_{HG}=0.083$ are shown with blue dashed and solid red histograms, respectively.}
\label{fig:partonangles}
\end{figure}

To further explore the effects of momentum-dependent new physics, consider the following effective couplings of the chromomagnetic dipole moment and Higgs-gluon kinetic operators: $c_{hgt}=-0.75$ and $c_{HG}=0.083$ at $\sqrt{s} = 8 ~ \rm{TeV}$ and  $c_{hgt}=-0.5$ and $c_{HG}=0.055$ at $\sqrt{s} = 14 ~ \rm{TeV}$ respectively. These points in parameter space are marked with stars in Fig.~\ref{fig:feynpph}. Much of our analysis focuses on these points in parameter space, though a wider set of parameters will be studied in Sec.~\ref{section:cmscutsleptonic}. These points are an ideal training ground for a targeted new-physics search strategy because their predicted $t \bar{t}h$ and Higgs production rates at $\sqrt{s} = 8 ~ \rm{TeV}$ and $\sqrt{s} = 14 ~ \rm{TeV}$ are nearly the same as those expected for the Standard Model despite the fact that $c_{HG}, c_{hgt} \ne 0$.  While the total $pp \to h$ and $pp \to t\bar t h$ cross sections for these points are SM-like, the $p_T$ distributions of the Higgs bosons is not, as shown in Fig.~\ref{fig:partonangles}. Accompanying this shift in the Higgs $p_T$ is a shift in the angular distributions of the $t, \bar{t}$ and $h$, which we show in the lower panels of Fig.~\ref{fig:partonangles}. A highly boosted Higgs could be balanced by just one of the tops, or the Higgs could recoil against the $t\bar t$ pair. The former configuration would be characterized by an imbalance in the two top momenta, while the latter is characterized by a decrease in the top and anti-top separation. The angular distributions in Fig.~\ref{fig:partonangles} indicate the latter scenario -- energetic Higgses recoiling against the top pair -- is more prevalent in the presence of $c_{HG}$ and $c_{hgt}$.

\section{Ferreting out high $p_T$ Higgs in same sign and trilepton $t \bar{t} h$ events}
\label{sec:dileptonhighpt}

In order to discriminate new top-Higgs distributions from Standard Model $t \bar{t}h$ events, we design and implement new variables aimed at the kinematic features in Fig.~\ref{fig:partonangles} -- namely that momentum-dependent new physics leads to a higher fraction of events with a high-$p_T$ Higgs recoiling against the top anti-top system. To exploit the differences shown in Fig.~\ref{fig:partonangles} , we must first recast the physics of these $p_T$ and angular distributions in terms of objects actually identified in the detector, i.e. jets, leptons, and missing energy. To do this, we need to specify the parton-level final state, which depends on how the Higgs, top, and anti-top decay.

We will focus primarily on $t\bar t h$ final states containing two or more leptons and at least one pair of same-sign leptons (SSDL). A separate discussion about the possibilities to study boosted Higgses in $t\bar th, h\to \gamma\gamma$ events will be presented later, in Sec.~\ref{section:cmscutsphotonic}. We do not study the possibilities for fermionic Higgs decays ($h \to \bar b b,\, h \to \tau^+\tau^-$), modes that are often used in jet-substructure based studies of boosted Higgses, both in the SM~\cite{Butterworth:2008iy,Plehn:2009rk} and beyond~\cite{Abdesselam:2010pt}, leaving these for future work. 

The CMS collaboration has searched for $t\bar t h$ events with at least two same charge isolated leptons, one or more b-jets passing the medium CSV working point, at least four hadronic jets~\cite{cmsmultilep}, and placed a 95\% confidence bound of $\mu_{t\bar{t}h} < 6.4$. For studies like this, cuts are made to find Higgs bosons  that decay through electroweak bosons  to at least one lepton ($h \to WW^*/ZZ^* \to \ell + X$) that are produced alongside a leptonic top or anti-top decay. The requirement of at least two same sign (same charge) leptons tends to exclude electroweak boson backgrounds such as $W^\pm/Z + \jets$ and $t\bar{t}$. The dominant backgrounds to same-sign lepton $t\bar t h$ events come from $t\bar{t}W^\pm + \jets$ (irreducible) and $t\bar t + \jets$ with one of the jets in the event faking a lepton.

With these preliminaries in mind, an obvious first place to look for high-$p_T$ Higgs is in the $p_T$ distributions of the Higgs decay products. For the same-sign dilepton and trilepton final states of interest here, we therefore look at the scalar sum over final state lepton transverse momenta,
\begin{align}
L_{p_T} = \sum_{\rm{isolated~leptons}} p_T^{\ell}.
\end{align}
In addition to being a proxy for a boosted Higgs, $L_{p_T}$ is a clean variable and does not require reconstructing the Higgs or top/anti-top. Reconstructing these heavy particles out of detector-level jets and leptons might get us closer to the distributions of Fig.~\ref{fig:partonangles}, but this procedure is fraught with combinatorial error. In addition to the $O(10)$ final state particles coming from the full $t\bar t h$ decay, reconstruction must contend with contamination from initial state radiation and potential ambiguities in how to partition the missing energy. For such complicated final states, these complications outweigh any benefits from reconstruction; we therefore seek variables like $L_{p_T}$ that do not require it.

Cutting on $L_{p_T}$ alone is a marginally effective discriminant, as we will show explicitly in Sec.~\ref{section:cmscutsleptonic}. While a cut requiring high $L_{p_T}$ is sensitive to events with boosted Higgs bosons, it is not ideal since, taken alone, high-$L_{p_T}$ favors events which happen to have many isolated leptons. The Standard Model $t \bar{t} h$ events are typically less collimated than events produced with new physics vertices we consider here, so they tend to contain more isolated leptons. Thus, SM $t\bar t h$ events often have a high $L_{p_T}$ from configurations with $3$ or more leptons and not because of a boosted Higgs. 

To further improve the search for the dimension-6 effective operators, we turn to the angular distributions shown in Fig.~\ref{fig:partonangles}. In circumstances where the Higgs boson balances against the net top anti-top system, the energy/transverse momentum per particle is higher on the Higgs side of the event (4 particles, as we consider $h \to WW^*, h \to ZZ^*$) than the $t\bar t$ side (6 particles). In a SSDL event, one lepton-neutrino pair comes from the Higgs side, while the other pair comes from the $t\bar t$ side. As the Higgs-side objects carry more energy, the $\slashed E_T$ vector will tend to point towards the Higgs side of the event and away from the top and anti-top. At the same time, all events are required to have at least one clearly tagged b-jet, which  provides a handle on the direction of the top or anti-top, especially at high-$p_{T,t},\, p_{T,\bar t}$. A large top-Higgs separation is therefore characterized by a large $b$-jet - $\slashed E_T$ separation. To capture this feature, we introduce the variable $B_{\phi}$, defined as the scalar sum over the $p_T$-weighted angular separation between b-tagged jets and the missing transverse energy vector,
\begin{align}
B_{\phi} = \sum_{\rm{b-tagged~ jets}} p_T^b \times |\Delta \phi_{\slashed{E_T},b}|.
\end{align}
Same-sign lepton events with a large top-Higgs separation are characterize by large $B_{\phi}$.  This variable has the added benefit that, when applied in conjunction with $L_{p_T}$, it also suppresses the $t\bar{t}W^\pm + \jets$ background. This is because, much like Standard Model $t \bar{t} h$ events, electroweak boson background events with a large $L_{p_T}$ tend to have a $\slashed{E}_T$ vector that falls more between the electroweak boson and the b-jets.

Requiring high values for the $B_\phi$ variable will select events in which the $b$-jets are distributed away from the missing transverse energy. While this will prefer events in which the Higgs is boosted against both top quarks, it nonetheless allows for standard model events in which the missing transverse energy vector happens to skew away from the $b$-tagged jet(s), which can happen in SM events with one top decaying leptonically near a Higgs boson (and the other highly boosted b-jet getting tagged). In order to exclude such events and further reinforce selection towards a high $p_T$ Higgs, we can define a variable parameterizing the angle between the leading lepton and the nearest b-jet. We find that using the lepton $p_T$ weighted separation of the leading $p_T$ lepton and closest b-jet in the $\Delta R$ plane,
\begin{align}
\Delta R_{\ell-b} =  p_T^\ell \times \Delta R_{\ell,b}
\end{align} 
is a complementary sieve for $B_\phi$ and $L_{p_T}$ in selecting highly-boosted Higgs events.

Finally, to fill out the array of variables sensitive to events with Higgs bosons boosted against tops, we define
\begin{align}
L_{\phi} = \sum_{\rm{isolated ~ leptons}} p_T^{\ell} \times |\Delta \phi_{\slashed{E_T},\ell}|.
\end{align}
which is the sum over isolated lepton $p_T$ multiplied by the angle between the lepton and missing transverse energy. Minimizing on this variable chooses events in which most of the lepton $p_T$ is sourced by a single highly boosted decay, which for our signal region is a high $p_T$ Higgs boson. This cut is complementary to requiring $b$ jets be removed from the missing energy vector (large $B_{\phi}$) -- instead, minimizing $L_\phi$ requires that the majority of the leptonic $p_T$ is near the missing energy vector. 

 One rather interesting result of this study, shown in Fig.~\ref{fig:histolsum}, is that the kinematic variable $L_\phi$ proves quite effective in eliminating Standard Model backgrounds like $t \bar{t}j$ + lepton fake. This is because the faked lepton $p_T$ of a same sign dielpton $t \bar{t}$ event will be uncorrelated with the $\slashed{E}_T$ vector, while the lepton $p_T$ of a $t \bar{t}h$ event tends to be closer to the $\slashed{E}_T$ vector. Thus the $L_\phi$ distribution of the background reaches $L_\phi$ values relatively unpopulated by either SM or new physics $t \bar{t}h$ events. We address this in more detail in Section \ref{section:cmscutsleptonic}.

\section{Using angular correlations to find boosted Higgs in $t \bar{t} h$ leptonic channels}
\label{section:cmscutsleptonic}

The angular variables defined in the preceding section were designed to select boosted Higgs events in $t \bar{t}h$ processes. Now we are ready to test out the variables in a Monte Carlo study. We implement the effective operators given in Section \ref{sec:ratemubounds} in FeynRules \cite{Alloul:2013bka,Alloul:2013naa} and generate events in MadGraph 5 \cite{oai:arXiv.org:1106.0522}. The parton-level events are showered and hadronized in Pythia \cite{oai:arXiv.org:0710.3820} then passed through a CMS-like detector via Delphes \cite{DELPHdeFavereau:2013fsa} to determine the number of expected LHC events for the cuts described hereafter. To incorporate higher-order QCD effects in the $t\bar t h$ signal, we multiply the next-to-leading order (NLO) cross sections stated in~\cite{Heinemeyer:2013tqa} (for $m_h = 126\, \gev$) by the ratio of the tree-level new physics $pp \to t\bar t h$ cross section to the SM tree-level $t\bar t h$ cross section\footnote{The tree-level cross sections in each scenario were calculated using the default MadGraph factorization/renormalization scheme and CTEQ6L1 parton distribution functions}. This procedure assumes that higher-order QCD~\cite{Beenakker:2001rj,Dawson:2002tg} effects are the same in the new physics scenario and in the SM. This is clearly an approximation, but we expect the search techniques we develop here are relatively unaffected, since any mismatch in the higher order effects between the SM alone and the SM augmented with new operators can be compensated by a shift in the range of new physics coefficients we are sensitive to (see Figure \ref{fig:feynpph}). For the backgrounds, we rescale the MadGraph-level cross sections to match the highest order calculations available, NLO~\cite{Campbell:2012dh} for $t\bar t + W$ and NNLO~\cite{Czakon:2013goa} for $t\bar t$. Most of the same-sign lepton background from $t\bar{t}$ comes from a jet faking a lepton within a $t \bar{t}$ event that has decayed to one or two real leptons. To generate this background, we followed the Madgraph-Pythia sequence outlined above, and then randomly selected 1\% of all jets in these events to receive a lepton charge with a random flavor and charge. After imposing the CMS-like cuts outlined below, we normalized the resulting $t \bar{t}$ sample to be proportional to our $t\bar t + W$ background, where this proportion was determined by the $t \bar{t}$/$t\bar t + W$ fraction found in \cite{cmsmultilep}. This rescaling factor is the ratio of the fake rate for  $t\bar t + \jets$ measured by CMS (assuming all ``non-prompt" events come from $t\bar t + \jets$) to the $1\%$ used in our simulation. When generating $14\,\tev$ events, we use this relative fake rate determined from $8\,\tev$ events to set the size of the $t\bar t + \jets$ background normalization. We neglect subdominant backgrounds, such as $t\bar t + Z/\gamma^*$, that make up ($\le 10\%$) of the total.

 Before using the new kinematic variables outlined in Section \ref{sec:dileptonhighpt}, we pass all events through the following, simplified cuts to mimic the event selection of \cite{cmsmultilep}.


\begin{itemize}
\item Leptons: Each event must have at least two leptons with $p_{T1} \geq 20 ~\rm{GeV}$, $p_{T2} \geq 10 ~\rm{GeV}$, and in the case of a third lepton, $p_{T2} \geq 7 ~\rm{GeV}$. Events must either have two same sign charge leptons or three or more leptons. Each lepton is isolated by the requirement that all charged and neutral particles in a cone of $\Delta R < 0.4$ have transverse momenta adding up to no greater than $0.4$ of the lepton's $p_T$.
\item Jets and b tags: There must be at least two jets, one of which must be b-tagged. We simulate the CMS Medium CSV working point b tag efficiency of 70\% identification and 1\% misidentification using a modified version of Delphes \cite{DELPHdeFavereau:2013fsa} and its CMS detector settings.
\item Each event must have a missing transverse energy of magnitude $\slashed{E_T} > 25 ~ \rm{GeV}$.
\end{itemize}

In addition to these standard cuts, we implement combinations of cuts on the angular variables defined in Section \ref{sec:dileptonhighpt}. The definition and utility of these kinematic variables are fleshed out in Section \ref{sec:dileptonhighpt}. Here we reiterate them for convenience.


\begin{itemize}
\item $L_{P_T}$, the scalar sum of isolated lepton transverse momentum.
\item $B_\phi$, the scalar sum of the angular ($\Delta \phi$) separations between the b-jets and $\slashed{E}_T$ weighted by the $b$-jet $p_T$.
\item $R_{\ell b}$, the angular separation ($\Delta R$) between the leading lepton and closest b-jet weighted by the $p_T$ of the lepton.
\item $L_{\phi}$, the scalar sum of $p_T^{\ell}$-weighted $\Delta \phi$ between the leptons and $\slashed{E}_T$. 
\end{itemize}

\begin{figure}[h!]
\centering
\begin{tabular}{cc}
\includegraphics[width=0.45\textwidth]{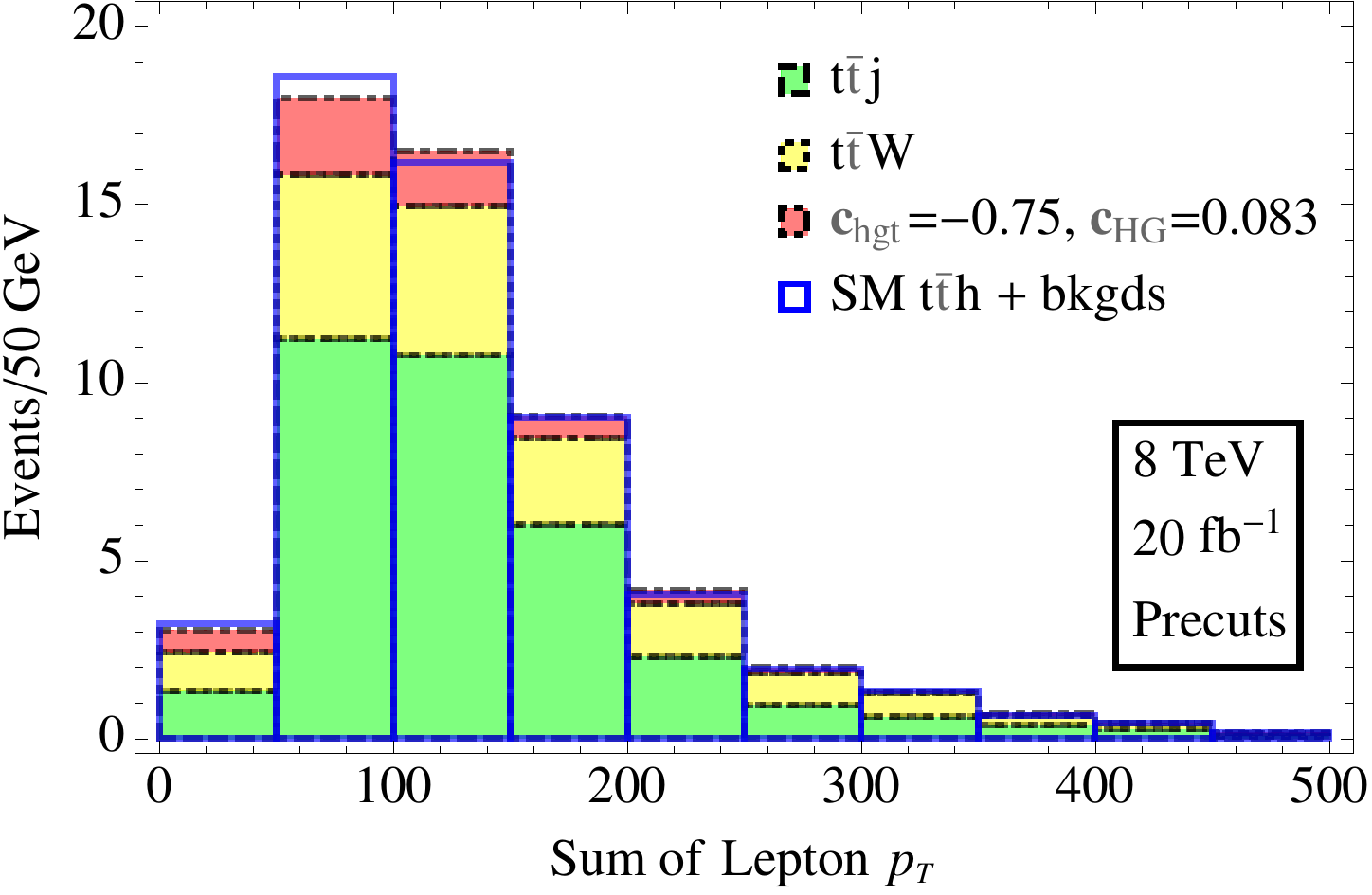}&\includegraphics[width=0.45\textwidth]{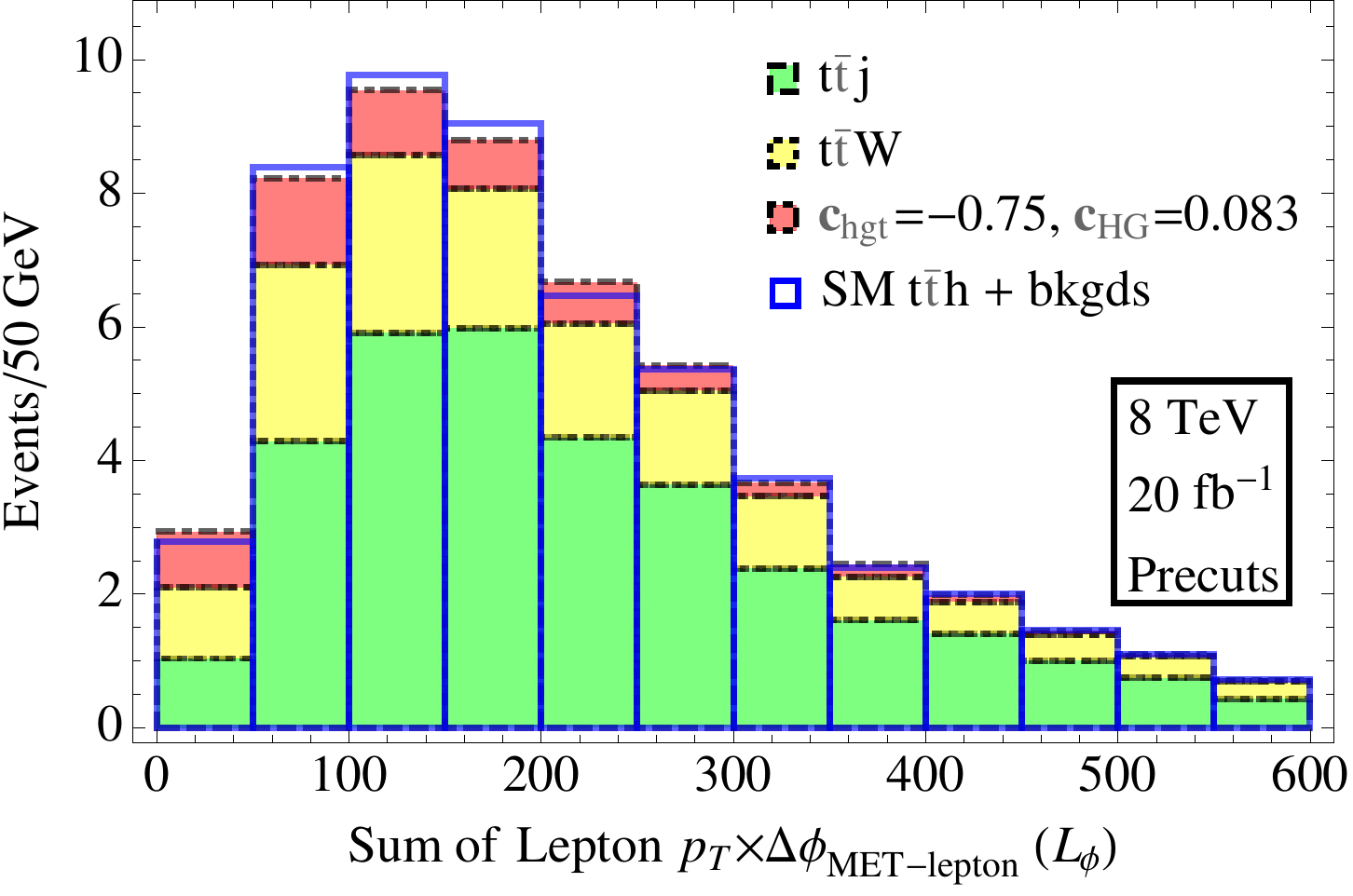} \\
\includegraphics[width=0.45\textwidth]{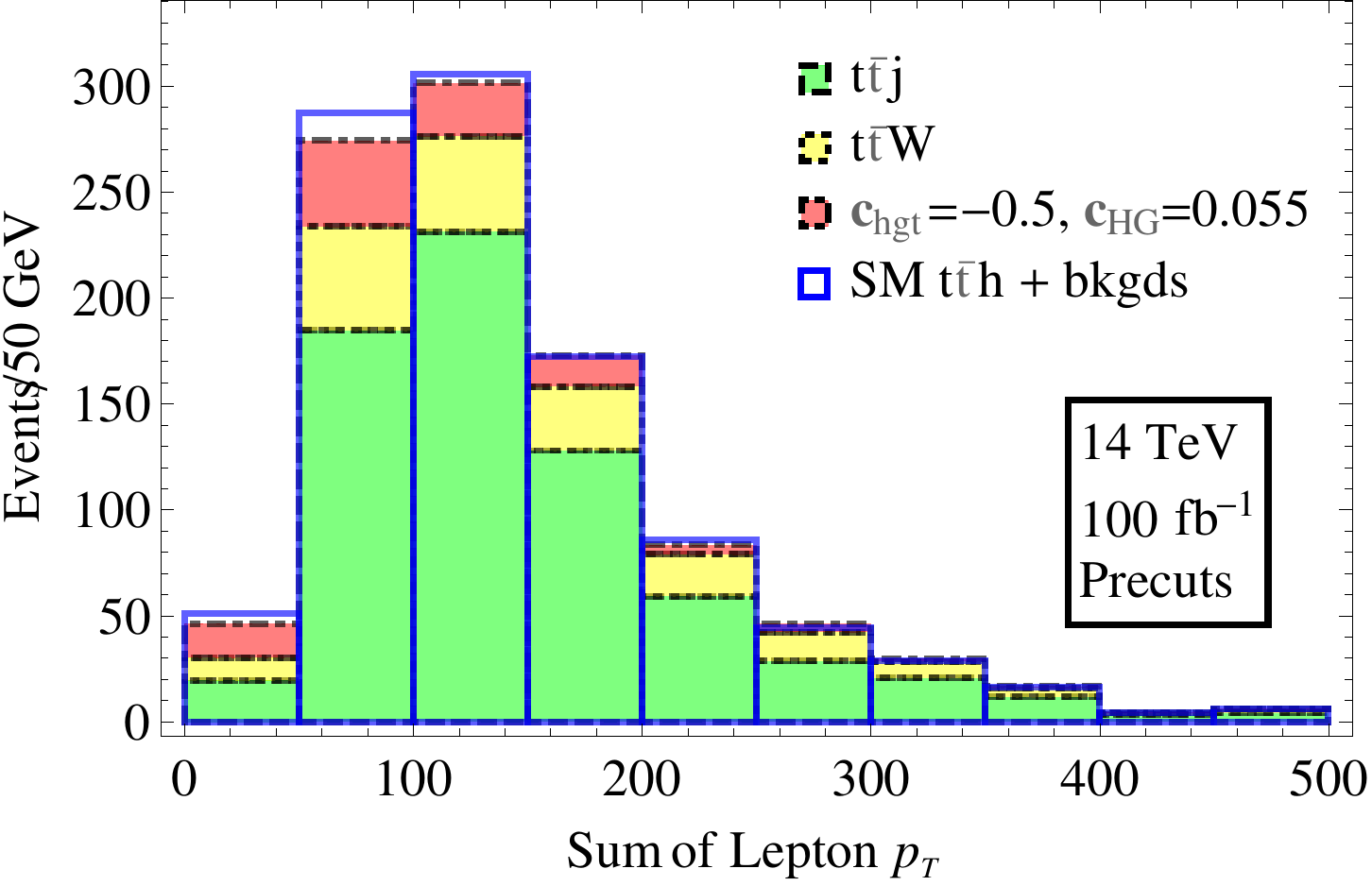}&\includegraphics[width=0.45\textwidth]{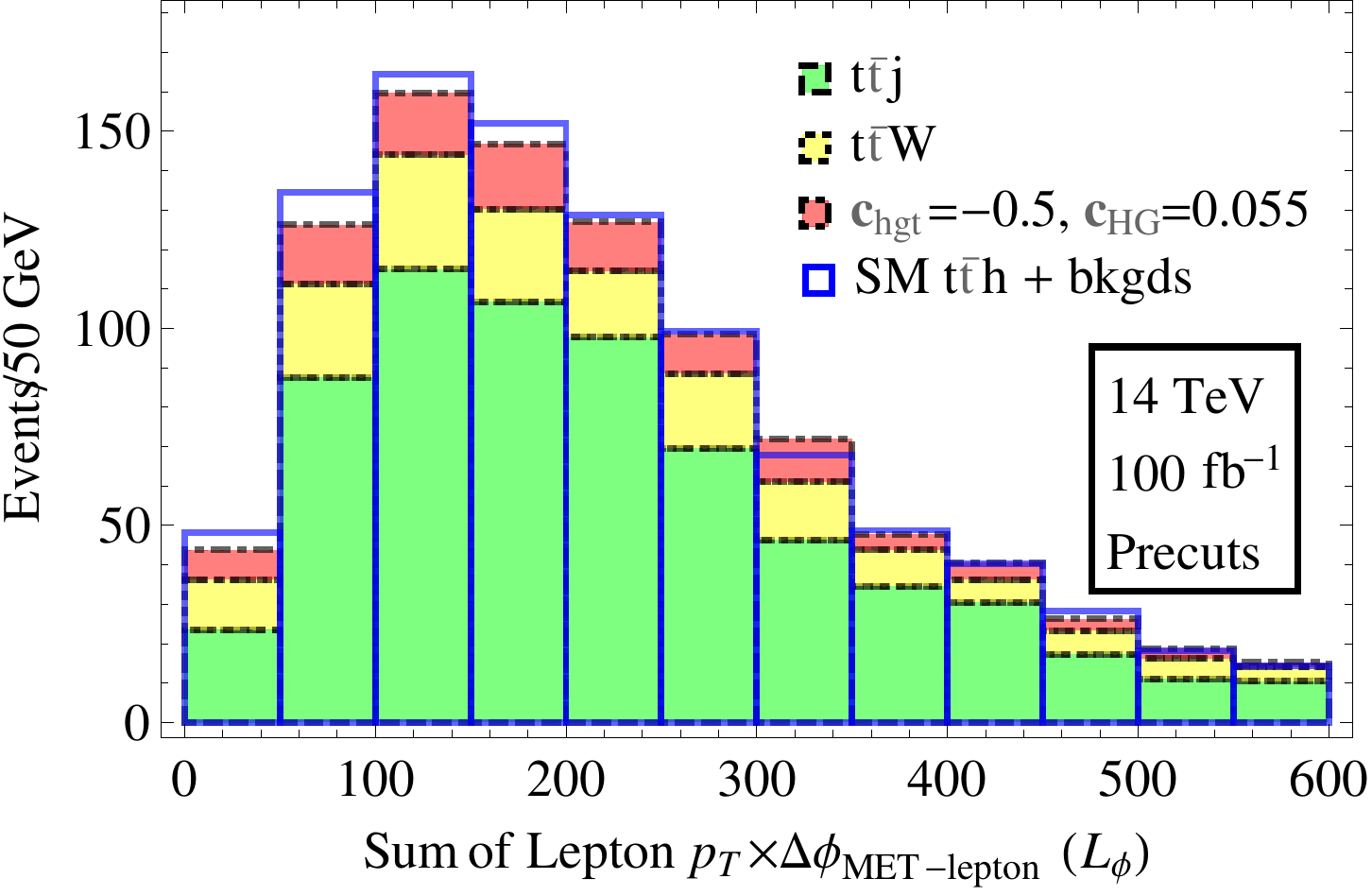} \\
\includegraphics[width=0.45\textwidth]{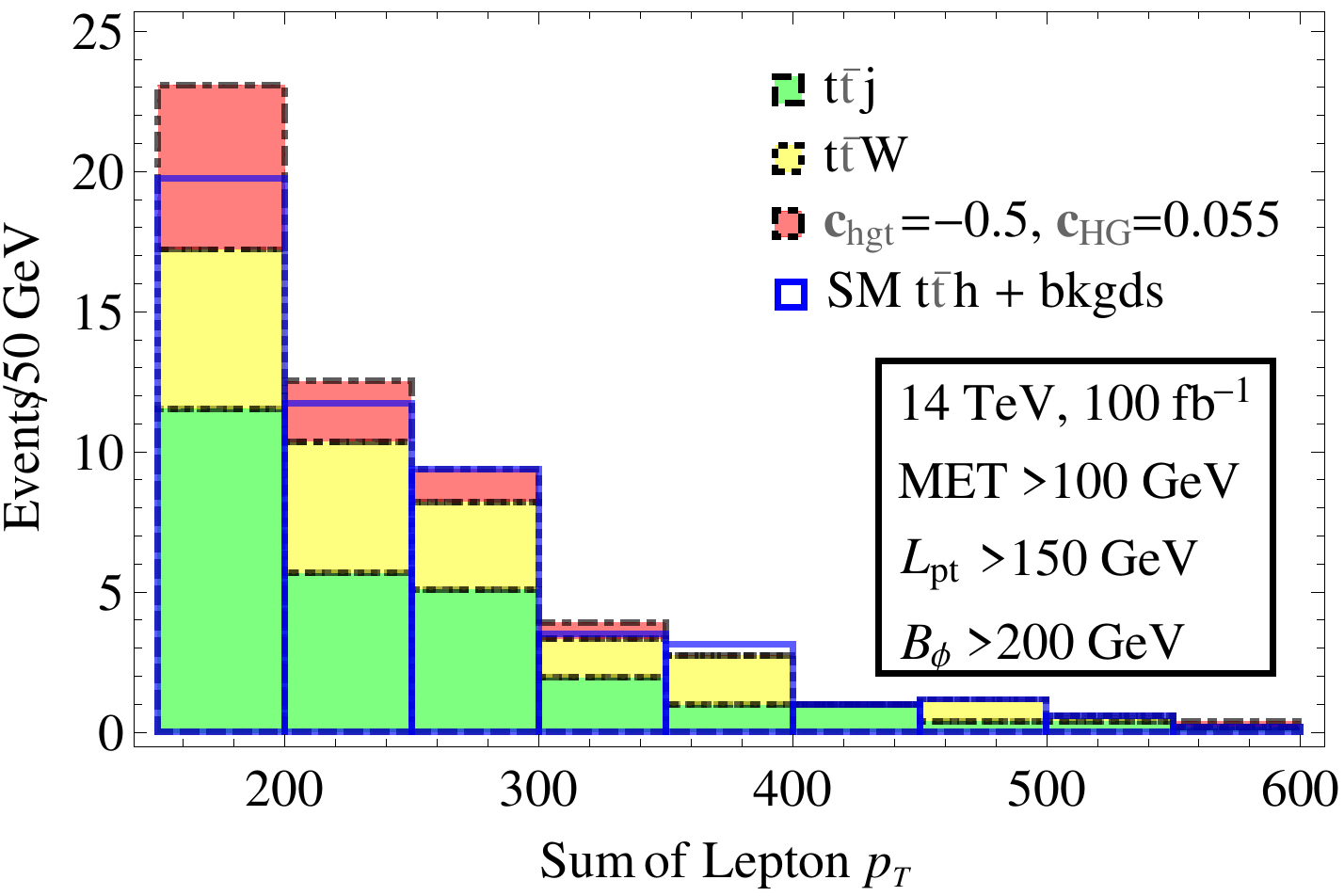} &\includegraphics[width=0.45\textwidth]{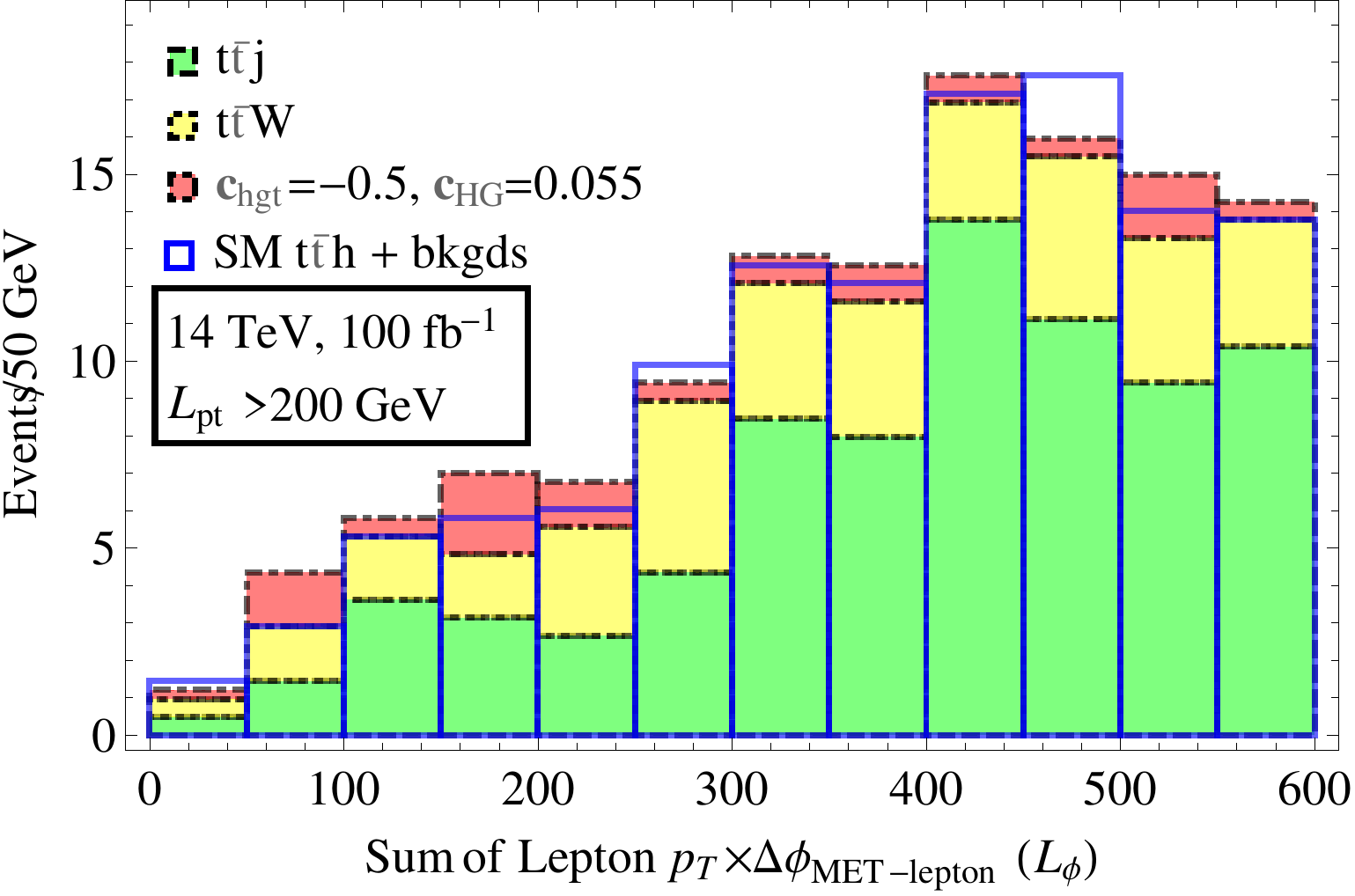}
\end{tabular}
\caption{The number of events surviving preliminary SSDL cuts and new angular variable cuts detailed in Section \ref{section:cmscutsleptonic} are displayed as a function of the scalar sum over lepton momenta ($L_{p_T}$) and the lepton $p_T$ weighted sum of transverse angular separation between $\slashed{E}_T$ and the lepton($L_\phi$). Events are binned for 8 and 14 TeV LHC runs with CMS-like precuts implemented as described in Section \ref{section:cmscutsleptonic}. The new physics events shown have dimension-6 couplings which render Standard Model and new physics cross-sections approximately equal, $\mu_{t \bar{t}h} \simeq 1$.}
\label{fig:histolsum}
\end{figure}

The four variables all target the same kinematic features -- a higher fraction of boosted Higgses, but they utilize different approaches. The variables $L_{p_T}$ and $L_{\phi}$ involve the lepton-$\slashed E_T$ system, while $R_{b-\ell}$ and $B_{\phi}$ measure the distance between a $b$ and the leptonic system. Similarly,  $L_{p_T}$ is a momentum sum, while the latter three variables are all $p_T$-weighted angular distributions. The $p_T$-weighted $\Delta \phi$ ($\Delta R$) distribution between two particles is similar to the transverse (invariant) mass $m_T$ of the system. Comparing e.g. $L_{p_T}$ with the transverse mass of the lepton-$\slashed E_T$ transverse mass
\begin{equation}
L_{\phi} = p_{T,\ell}\, \Delta \phi_{\slashed E_T -\ell} \quad \text{vs.} \quad m_{T,\slashed E_T-\ell}^2 = 2\,\slashed E_T\, p_{T,\ell}\,(1-\cos(\Delta \phi_{\slashed E_T-\ell}) ) \sim \slashed E_T\,p_T\,\Delta\phi^2_{\slashed E_T -\ell},
\end{equation}
the latter has an extra factor of $\slashed E_T$ and $\Delta \phi$. The trends in $L_{\phi}$ and $m_{T,\slashed E_T-\ell}$ will be the same, however the correlations between these variables and $\slashed E_T$ are different. We find working with $L_{\phi}$ and $\slashed E_T$ separately, rather than in the package $m_T$ is more helpful -- though a study incorporating both types of variables would be interesting.

The effect of a few cut combinations on the two $(c_{HG}, c_{hgt})$ benchmark points are shown in Fig.~\ref{fig:histolsum}. The backgrounds $t\bar t + \jets$ and $t\bar t + W^{\pm}$ sit at the bottom of the histograms, followed by the $t\bar t h$ events for the benchmark new physics scenarios. The result of stacking the SM $t\bar t h$ signal atop the background is overlaid for comparison. The top two (middle two) panels show the $L_{p_T}$ and $L_{\phi}$ at $8\,\tev,\, 20\, \fbinv$ ($14\,\tev,\, 100\,\fbinv$) with only the basic CMS cuts. The trend towards higher $p_T$ Higgs for the new physics events can already be seen, as the SM distributions are slightly higher for $p_{T,\ell} \le 100\,\gev, L_{\phi} \le 250\,\gev$. While neither $L_{p_T}$ nor $L_{\phi}$ alone are ideally suited for distinguishing new momentum-dependent physics from the SM, the combination of high $L_{p_T}$ and low $L_{\phi}$ does work well. We can see this in the bottom left panel, where we show $L_{\phi}$ after cutting $L_{p_T} > 200\,\gev$ on $14\,\tev$ events; the new-physics events are bunched at $L_{\phi} \le 250\,\gev$. A similar separation between SM and new-physics events is also visible in the bottom right panel, where the $L_{p_T}$ cut has been lowered but supplemented with cuts on $B_{\phi}$ and $\slashed E_T$.

In Table \ref{tab:multilep} we apply several cut combinations, including those shown in Fig.~\ref{fig:histolsum}, to the benchmark $(c_{HG}, c_{hgt})$ points, comparing with the number of SM $t\bar t h$ events and the number of SM background events. The cut combinations we show are not intended to be exhaustive, nor have they been optimized for either of the benchmarks. Instead, our goal is to point out several different variables that can be used -- either alone or in combinations -- to target excess high-$p_T$ Higgses. Having multiple variables is useful for a number of reasons; the different variables will have different systematic uncertainties, and correlations among multiple variables can be exploited by multi-variate analysis (MVA) to further enhance discriminating power. We emphasize that while the absolute numbers of events (both for the signals and the background) are subject to approximations in our detector tools and in how we implement the cuts, the relative rates (i.e. new physics vs. SM) are robust.

\begin{table}[h!]
\begin{center}
\footnotesize
\begin{tabular}{c}
\begin{tabular}{|l|r|r|r|r|}
\hline
$\sqrt{s} = $8 TeV, 19.6 fb$^{-1}$, $c_{hgt}=-0.75$, $c_{HG} = 0.083$& $t \bar{t} + jets$ & $t \bar{t}W^\pm + jets$ & SM Signal  & NP Signal \\
\hline
\hline
SSDL preliminary cuts only & 34 & 16 & 5.8 & 5.6 \\
$L_\phi < 250$  & 21 & 10 & 4.9 & 4.4 \\
$L_{p_T} > 200$  & 5.0 & 3.7 & 0.40 & 0.73  \\
$L_\phi < 150$, $B_\phi > 150$, $\Delta R_{\ell b}>150$  & 2.1 & 1.3 & 0.45 & 0.69 \\
$\slashed{E}_T > 100$, $L_{p_T} > 100$, $L_\phi < 150$& 0.59 & 0.89 & 0.25 & 0.46\\
\hline
\end{tabular}
\\
\\
\begin{tabular}{|l|r|r|r|r|}
\hline
$\sqrt{s} = $14 TeV, 100 fb$^{-1}$, $c_{hgt}=-0.5$, $c_{HG} = 0.055$& $t \bar{t} + jets$ & $t \bar{t}W^\pm + jets$ & SM Signal  & NP Signal \\
\hline

\hline
SSDL preliminary cuts only & 697 & 185 & 128 &  114 \\
$L_\phi < 250$  & 428 & 107 & 106 & 91 \\
$L_{p_T} > 200$  & 124 & 51 & 8 & 13  \\
$L_{p_T} > 200$, $L_\phi < 250$ & 10.9 & 8.1 & 2.3 & 4.1\\
$\slashed{E}_T > 80$, $L_\phi < 100$, $B_\phi > 150$, $\Delta R_{\ell b}>150$  & 8.1 &  5.9 & 4.5 & 6.0 \\
$\slashed{E}_T > 100$, $L_{p_T} > 150$, $B_\phi > 200$    & 28 & 18 & 4.8 & 10  \\
\hline
\end{tabular}
\\
\end{tabular}
\caption{Expected events after cuts for 19.6 $\rm{fb}^{-1}$ LHC luminosity taken at 8 TeV and 100 $\rm{fb}^{-1}$ at 14 TeV. All rows displaying cuts incorporate the simplified same sign dilepton (SSDL) cuts detailed in Section \ref{section:cmscutsleptonic}. Cuts are given in units of GeV for kinematic variables defined in Section \ref{sec:dileptonhighpt} and listed in Section \ref{section:cmscutsleptonic}. Note that for the values of $c_{hgt}$ and $c_{HG}$ indicated, Standard Model $t \bar{t} h$ rates match new physics rates ($\mu_{t\bar{t}h} \simeq 1$).} 
\label{tab:multilep}
\end{center}
\end{table}
The second and third lines in table~\ref{tab:multilep} verify that cuts on the sum of lepton $p_T$ or $L_{\phi}$ alone are not as successful at picking out new physics events as a variety of cuts on angular final state distributions. Particularly, a 100 GeV cut on missing energy combined with requirement of at least 150 GeV lepton $p_T$ and $B_\phi > 200 ~ \rm{GeV}$ is highly effective in selecting for boosted Higgs boson events while reducing SM backgrounds. Indeed, the fourth and fifth rows of of 8 and 14 TeV tables illustrate that very effective cuts can be designed without directly cutting on the sum of lepton $p_T$. Surveying the final few rows of each table, it is apparent that there is a trade-off between reducing SM background to $t \bar{t}h$, which tend to diminish with more rigorous cuts on $L_\phi$, and increasing the new-physics/SM ratio by cutting on $B_\phi$ and $\Delta R_{\ell b}$. The results for twenty inverse femtobarns of events at eight TeV indicate that a fruitful study of already existing data could be conducted which would limit or possibly detect new Higgs couplings via Higgs kinematics in $t \bar{t}h$ final states. 

\begin{figure}[h!]
\centering
\begin{tabular}{cc}
\includegraphics[width=0.45\textwidth]{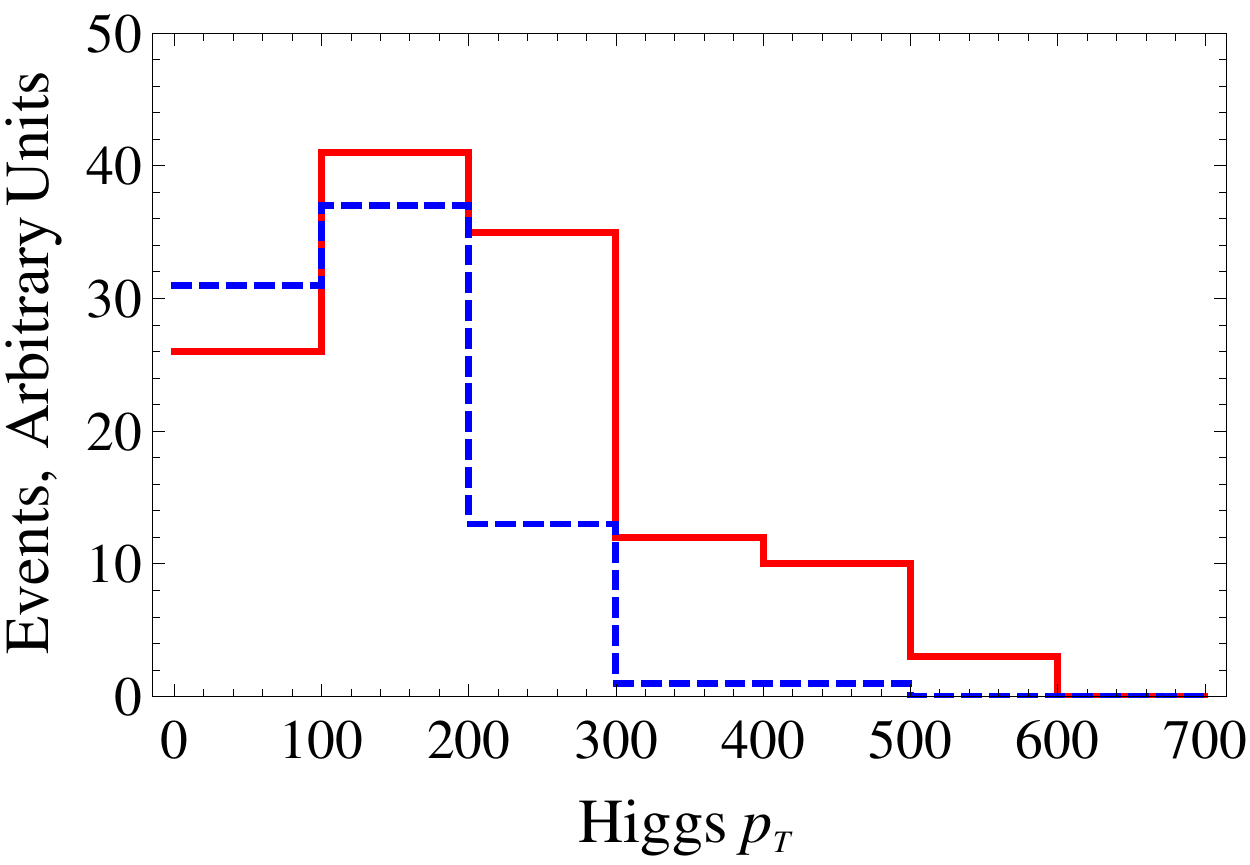}&\includegraphics[width=0.45\textwidth]{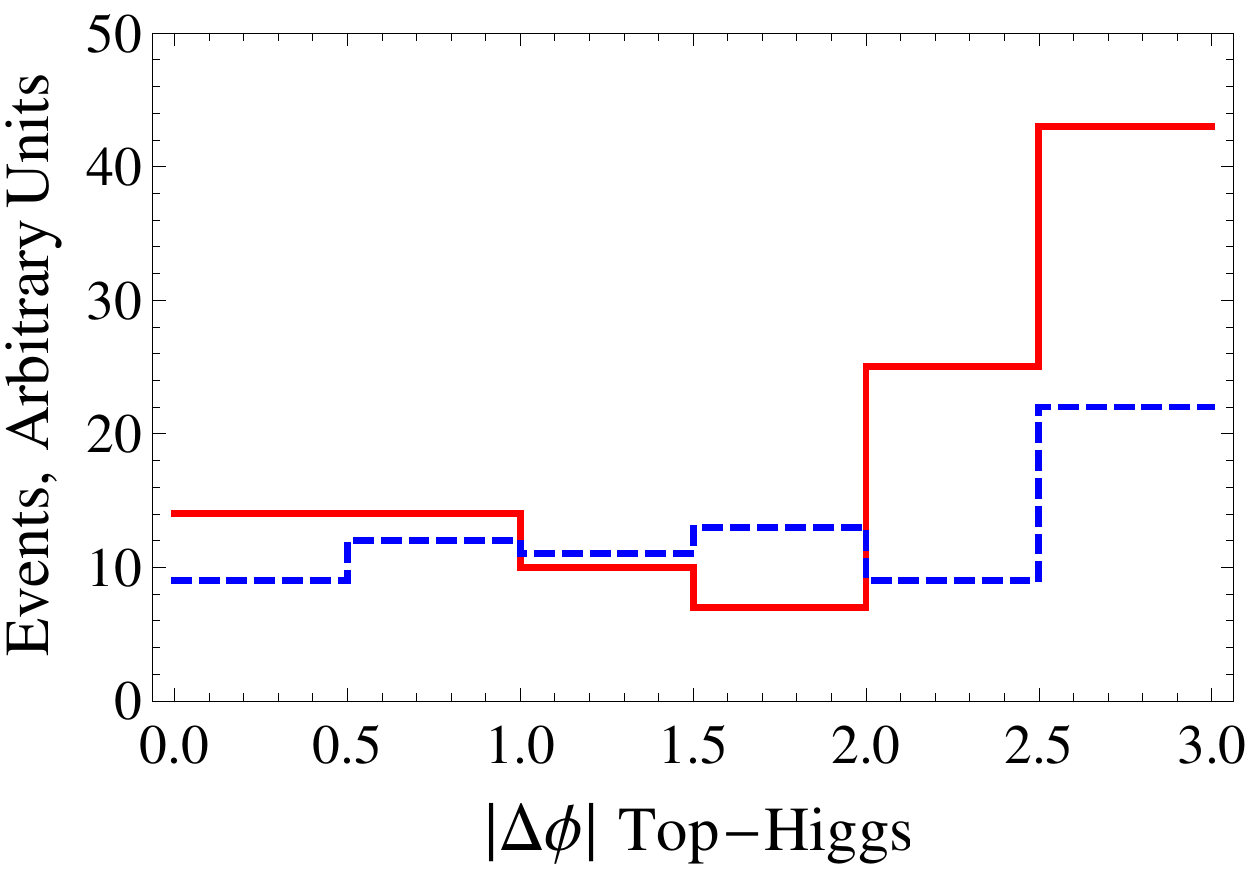} \\\includegraphics[width=0.45\textwidth]{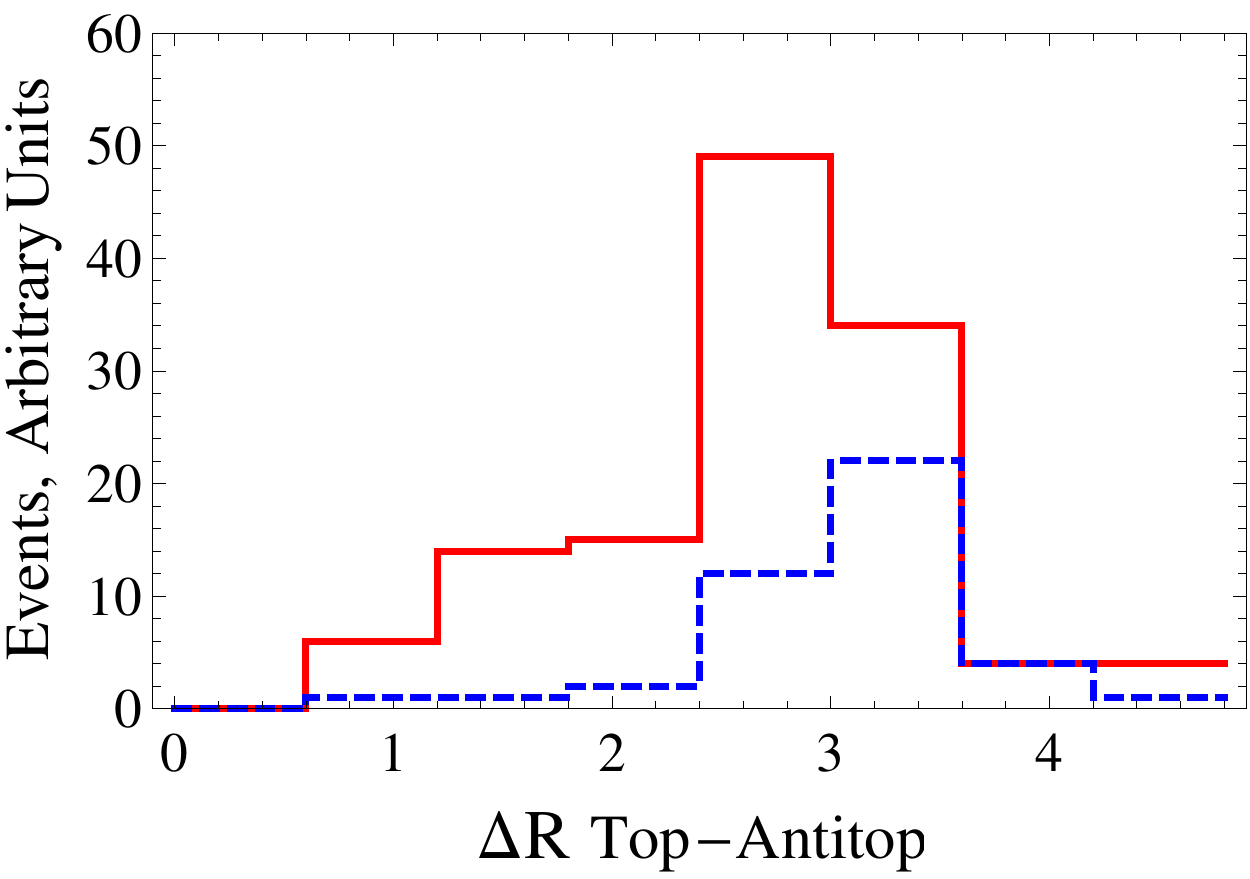} &\includegraphics[width=0.45\textwidth]{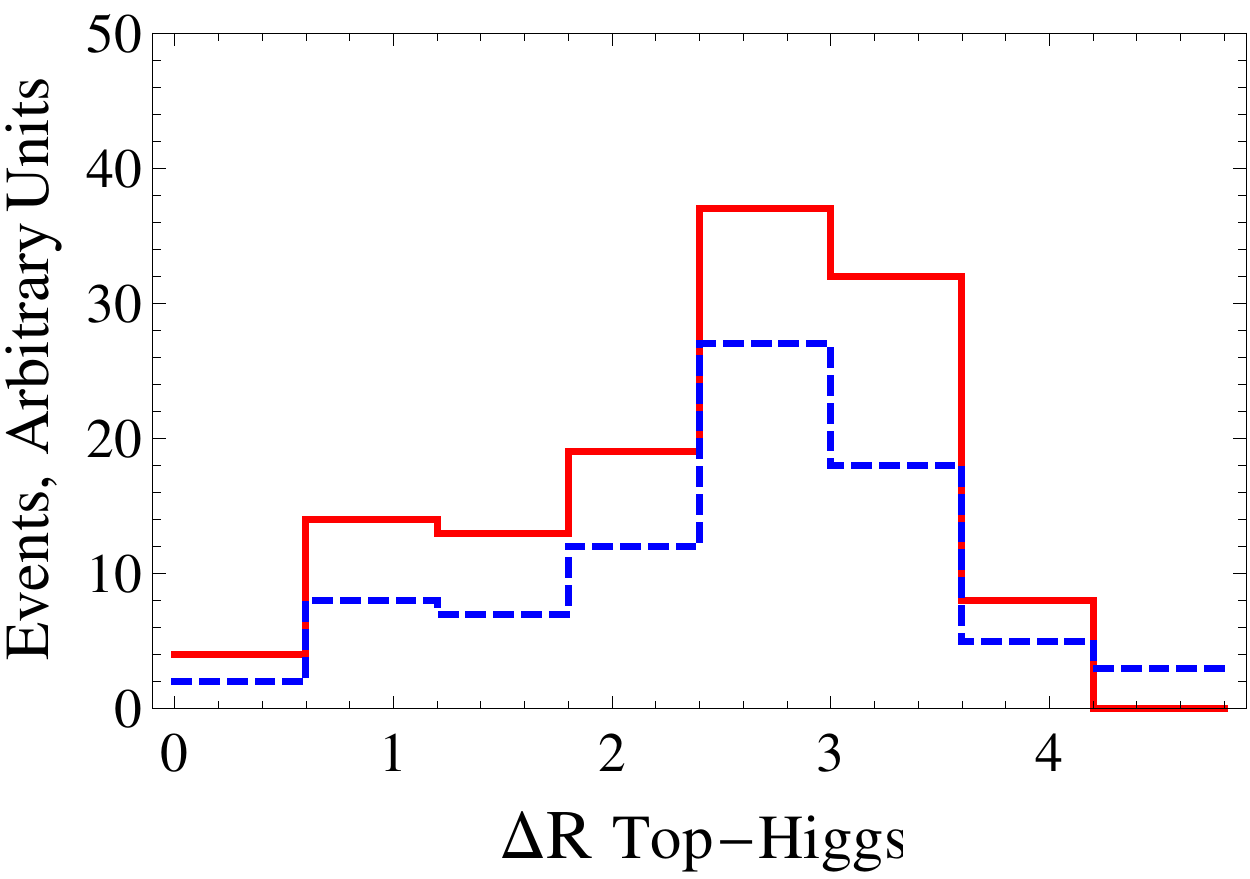}
\end{tabular}
\caption{Kinematic distributions after cuts: The initial state $p_T$ of Higgs bosons and angular separation of Higgs and top quarks are plotted for 8 TeV events which remain after cuts on angular variables detailed in the text, $\Delta R_{\ell b},B_{phi} > 150 ~\rm{GeV},$ and $L_{\phi} <  150 ~\rm{GeV}$. As before, the Standard Model events passing these cuts are displayed with a blue dashed line, and the new physics events with operator coefficients $c_{hgt}=-0.75$ and $c_{HG}=0.083$ are binned with red solid lines. The number of events plotted is in arbitrary units, though the ratio of new physics to SM events accurately reflects the ratio resulting from the kinematic cuts quoted.}
\label{fig:partonkinaftercuts}
\end{figure}

To cross-check the intuition behind our proposed variables, in Fig.~\ref{fig:partonkinaftercuts} we plot the parton ($t\bar t h$-level) distributions from Fig.~\ref{fig:partonangles} for events that survive the cut combination $\Delta R_{\ell b},B_{\phi} > 150 ~\rm{GeV},$ and $L_{\phi} <  150 ~\rm{GeV}$ (the fourth row of the top part of table~\ref{tab:multilep}). The ratio of new physics to SM events accurately reflects the affect of the cuts, but the total number of events shown is arbitrary. The post-cut distributions clearly verify the arguments in Sec.~\ref{sec:dileptonhighpt}; the surviving events are characterized by large Higgs $p_T$ and a large separation, both in $\Delta \phi$ and $\Delta R$ between the Higgs and top $t\bar t$ system. Notice that the Higgs $p_T$, while large, are still small compared to the EFT cutoff $\Lambda$, so our results do not rely on pushing the theory into unreliable territory.
 
While Table \ref{tab:multilep} shows the effectiveness of an angular kinematic search for boosted Higgs in Higgs-top events, it is useful to remember that the benchmark new physics parameters shown above produce event rates that are conservatively low -- the parameters were chosen to give $t \bar{t}h$ rates {\em equal} to Standard Model event rates, while $\mu_{t\bar t h}$ as high as $6.4$ are still consistent with current data. Thus, when considering new physics points with a non-SM $t\bar t h$ rate in addition to non-SM Higgs $p_T$ distribution, the discovery reach of these cuts extends farther than Table \ref{tab:multilep} demonstrates. In Table \ref{tab:multilep_several}, we apply these cuts to a menagerie of model parameters, and find that these cuts do better than simply gaining a linear increase in boosted Higgs events for a linear increase in new physics $t \bar{t} h$ production cross-section. Indeed, for the model point $c_{hgt}=-0.75$, $c_{HG}=0.083$, which at 14 TeV has a cross-section 1.5 times the SM, we find that these angular cuts select five times as many  events as in the SM. Even when comparing two different new physics points, for example  $c_{hgt}=-0.75$, $c_{HG}=0.083, \mu_{t\bar t h} = 1.5$ and $c_{hgt}=-0.5$, $c_{HG}=0.055,\ \mu_{t\bar t h} = 1.0$, the increase in event yield after cuts  is greater than the change in $\mu_{t\bar{t}h}$. 

\begin{table}[h!]
\begin{center}
\begin{tabular}{c}
\begin{tabular}{|c|c|c|}
\hline
$\sqrt{s} = $14 TeV, 100 fb$^{-1}$, & $\mu_{t\bar{t}h}$ & events after   \\
parameter points listed & &$\slashed{E}_T > 100$,  \\
&&$L_{p_T} > 150$,\\
&&$B_\phi > 200$
\\
\hline

\hline
Standard Model & 1 &  4.8 \\
$c_{hgt}=-0.75$, $c_{HG} = 0.083$ & 1.5 &  24 \\
$c_{hgt}=-0.75$, $c_{HG} = -0.13$ & 1.4 & 15 \\
$c_{hgt}=-0.5$, $c_{HG} = 0.055$ & 0.99 & 10  \\
$c_{hgt}=-0.5$, $c_{HG} = -0.16$ & 0.95 & 8.2\\
$c_{hgt}=0.25$, $c_{HG} = -0.028$ & 1.2 & 12  \\
\hline
\end{tabular}
\\
\\
\begin{tabular}{|c|c|c|}
\hline
$\sqrt{s} = $8 TeV, 20 fb$^{-1}$, & $\mu_{t\bar{t}h}$ & events after   \\
parameter points listed & &$L_{\phi} < 150$,  \\
&&$B_\phi > 150$,\\
&&$\Delta R_{\ell b} > 150 $
\\
\hline

\hline
Standard Model & 1 &  0.45 \\
$c_{hgt}=0.5$, $c_{HG} = -0.27$ & 2.1 &  4.1 \\
$c_{hgt}=-1.75$, $c_{HG} = -0.023$ & 3.3 & 4 \\
$c_{hgt}=1$, $c_{HG} = -0.32$ & 4.6 & 8.2  \\
\hline
\end{tabular}
\end{tabular}
\caption{In these tables we demonstrate new physics detection prospects for boosted Higgs in $t \bar{t}h$ by showing signal events expected for 20 fb$^{-1}$ at an 8 TeV LHC and 100 fb$^{-1}$ ar 14 TeV in parameter space with $\mu_{t \bar{t}h} \neq 1$. Note that the SM backgrounds for these cuts are shown in Table \ref{tab:multilep}.}
\label{tab:multilep_several}
\end{center}
\end{table}

Finally, let us return to the second row of table~\ref{tab:multilep}, where we have shown the effect of imposing a minimal cut on $L_\phi$. While this cut alone lessens the ratio of new physics to Standard Model $t \bar{t}h$, a cursory examination will reveal that the SM $t\bar{t}W^{\pm}$ and $t \bar{t}+\jets$ backgrounds are halved for a modest decrease in SM signal events. Thus, we propose this angular variable could enhance existing SM $t \bar{t}h$ studies.

\section{Seeking non-standard $t \bar{t} h$ kinematics in the digamma decay channel}
\label{section:cmscutsphotonic}

The less common, but more easily reconstructed Higgs decay to two photons provides another possible detection channel for digging out boosted Higgs bosons produced with $t \bar{t}$ pairs. Studies of this detection channel have been conducted using backgrounds estimated with control regions \cite{cmsdiphoton} or fitted to simple shapes~\cite{atlasdiphoton, ATLAS-CONF-2013-012}. Actually, because the Higgs branching ratio to digamma is relatively small, and $pp \rightarrow t \bar{t} h$ studies need to impose stringent cuts to exclude hadronic backgrounds, studies of boosted Higgs in this channel will have to employ minimal cuts in order to keep any events at all. More to the point, the Higgs boson is fully reconstructible in the $\gamma\gamma$ decay channel, so we do not need to resort to angular correlations; after imposing a $m_{\gamma\gamma}$ window to filter out photon pairs that could not come from the Higgs, we study Higgs kinematics using the scalar sum over photon $p_T$,
\begin{align}
\Gamma_{p_T} = \sum_{\rm{isolated~photons}} p_T^{\gamma}.
\end{align}

Studies of diphoton $t \bar{t}h$ events, $t \bar{t} h \rightarrow t \bar{t} \gamma \gamma $ employ control regions with single isolated photons and non-isolated photons to determine the background for Standard Model production of $ t \bar{t} h, h \rightarrow \gamma \gamma$. While a control region analysis is beyond the scope of this work, here we demonstrate that cutting on the sum of photon $p_T$ is useful for distinguishing SM diphoton $t \bar{t}h$ events from those produced by new physics giving a large fraction of high $p_T$ Higgs. The cuts we employ for SM signal and new physics signal follow those of \cite{cmsdiphoton,atlasdiphoton} 
\begin{itemize}
\item For events in the hadronic channel: Two photons, one with $p_T$ greater than 25 GeV, another with $p_T$ greater than 33 GeV. At least five identified jets, one of which must have a Medium CSV b-tag, which we simulate with Delphes by assigning a b-tag efficiency of 70\% and a tau or charm quark mistag efficiency of 1.5\%.
\item For events in the leptonic channel: Two photons, one with $p_T$ greater than 25 GeV, another with $p_T$ greater than 33 GeV. At least two identified jets, one of which must have a Medium CSV b-tag, which we simulate with Delphes as described in the hadronic channel.
\end{itemize}

\begin{figure}[h!]
\centering
\begin{tabular}{cc}
\includegraphics[scale=.7]{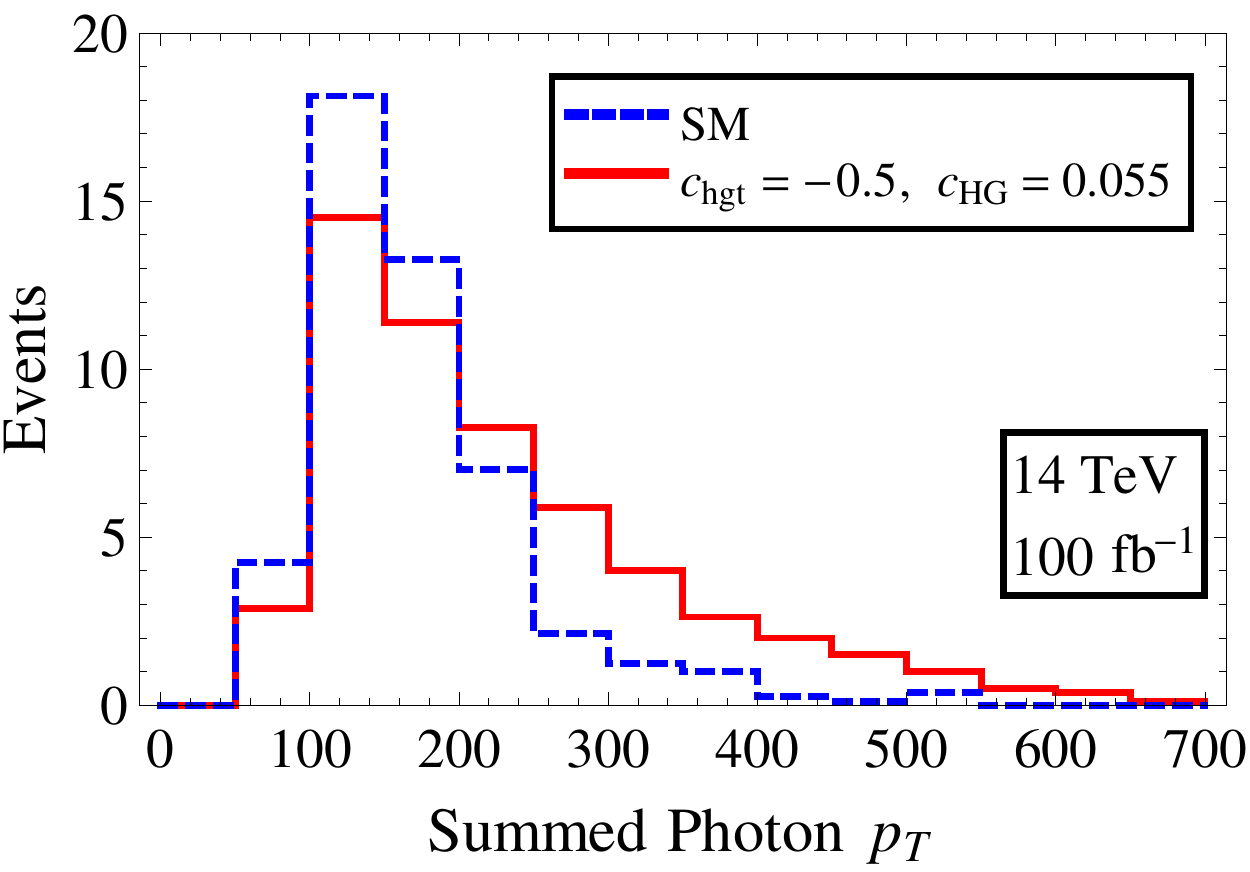}
\end{tabular}
\label{fig:phot}
\caption{Simulated 14 TeV LHC diphoton $t \bar{t}h$ events are shown for a luminosity of 100 fb$^{-1}$. Events are binned over scalar sums of diphoton $p_T$ for both Standard Model events and new physics events with couplings $c_{HG}=-0.5$ and  $c_{hgt}=0.055$.}
\end{figure}

Using simulated LHC events generated in the manner described in Section \ref{section:cmscutsleptonic}, we can determine whether this event selection is sensitive to a boosted $p_T$ Higgs. Figure \ref{fig:phot} displays Standard Model and new physics coupled ($c_{HG}=-0.5,~ c_{hgt}=0.055$) events. These two sets of events, whose cross-sections are essentially equal, are clearly separable with cuts on a scalar sum of photon $p_T$. Such a survey is limited by statistics -- with 100 fb$^{-1}$, only $\sim 10$ events remain in the new physics case -- while the SM background is potentially an order of magnitude larger~\cite{cmsdiphoton,atlasdiphoton}.

\section{Conclusions}
\label{sec:conclusion}
In this work we have developed a set of simple, reconstruction-free angular variables that are sensitive to $t\bar t h$ events with non-standard kinematics sourced by the dimension-6 Higgs-gluon kinetic ($c_{HG}$) and chromomagnetic dipole ($c_{hgt}$) operators. These are the two lowest-dimension operators which contribute to a boosted $p_T$ Higgs spectrum in $t \bar{t}h$ final states, and we have implemented  angular variable event selection strategies and shown detection prospects for the 8 TeV LHC with 20 fb$^{-1}$ and a 14 TeV LHC with 100  fb$^{-1}$. 

While other dimension-($n \geq 5$) operators could alter the production of, e.g., the Higgs, $t \bar{t}$, and $t \bar{t} h$, we have shown that the unique momentum structure of the two operators studied here provides a distinct and independent avenue for finding novel QCD or Higgs dynamics. We have focused particular attention on building upon prior successful $t \bar{t}h$ searches \cite{cmsmultilep} in the same sign lepton channel. We find that sums over b-jet and lepton $p_T$ weighted angles between these and the missing transverse energy of an event preferentially select boosted $p_T$ Higgs events. It has been shown explicitly how this detection regime complements kinematic-blind event rate studies of $pp \to h$, $t \bar{t}$, and $t \bar{t} h$; values of $c_{HG},c_{hgt}$ which predict the same number of $t \bar{t} h$ events as the Standard Model can nevertheless be selected for in $>2:1$ ratio to the SM using the angular variables detailed in Section \ref{sec:dileptonhighpt}. 

We have also briefly studied the detection prospects for of a boosted top-Higgs system in the diphoton channel. While this channel is limited by much lower statistics, a simple scalar sum over photon $p_T$ proves a useful discriminant of SM versus boosted events.

Finally, in addition to providing a kinematic portal on non-standard top-Higgs kinematics, the angular variables devised in this study also show promise as tools to improve ongoing LHC studies of $t \bar{t} h$. Specifically, the  sum over lepton $p_T$ weighted lepton-$\slashed E_T$ separation ($L_\phi$) is shown to significantly reduce $t \bar{t} W^{\pm}$ and $t \bar{t} +$ lepton fakes,  the dominant backgrounds for multilepton $t\bar t h$ studies, while only slightly reducing  the number of signal events, even assuming a Standard Model Higgs $p_T$ distribution. \\

{\bf Acknowledgements}
We wish to thank Andrew Brinkerhoff, Ahmed Ismail, Kevin Lannon, Veronica Sanz, Shufang Su, and Anna Woodard for useful discussions. The work of AD was partially supported by the National Science Foundation under grant PHY-1215979.

%

\end{document}